\documentclass[twocolumn]{aastex62}
\usepackage[normalem]{ulem}
\usepackage{verbatim}
\usepackage{graphicx} 
\usepackage{dcolumn}
\usepackage{bm}
\usepackage{xcolor}
\usepackage{url}
\usepackage{amsmath}
\usepackage{morefloats}
\usepackage{times}
\usepackage[varg]{txfonts}
\usepackage{multirow}
\usepackage{lineno}
\usepackage{hyperref}
\usepackage{notoccite}
\usepackage{acronym}
\usepackage{xspace}
\usepackage[utf8]{inputenc}

\usepackage{color}

\definecolor{scarlet}{rgb}{1.0, 0.13, 0.0}

\def\ie{{i.e.~}}
\def\eg{{e.g.~}}

\begin{document}

\newcommand{\kmsMpc}{\ensuremath{\mbox{km s}^{-1} \,\mbox{Mpc}^{-1}}}
\newcommand{\Mpc}{\ensuremath{\mbox{Mpc}}\xspace}
\newcommand{\sqdeg}{\ensuremath{\text{deg}^2}\xspace}

\newcommand{\HubbleMeasCombinedCounterpart}{\ensuremath{69.6^{+20.4}_{-8.6}}} 
\newcommand{\HubbleMeasCombinedCounterpartFLATLOG}{\ensuremath{68.7^{+17.0}_{-7.8}}} 
\newcommand{\HubbleMeasGWOneSevenZeroEightOneSevenCounterpartFLATLOG}{\ensuremath{68.7^{+17.5}_{-8.3}}} 
\newcommand{\HubbleMeasCombinedCounterpartFLATLOGMEDIAN}{\ensuremath{74.9^{+39.2}_{-13.9}}} 

\newcommand{\closest}{\ensuremath{320^{+120}_{-110}}\xspace}
\newcommand{\farthest}{\ensuremath{2840^{+1400}_{-1360}}\xspace}
\newcommand{\mosttight}{\ensuremath{39}\xspace}
\newcommand{\leasttight}{\ensuremath{1666}\xspace}
\newcommand{\bnsdistance}{\ensuremath{40^{+10}_{-10}}\xspace}

\newcommand{\HubblePriorMin}{\ensuremath{20}\xspace}
\newcommand{\HubblePriorMax}{\ensuremath{140}}

\newcommand{\OmegaMatter}{\ensuremath{0.308}\xspace}
\newcommand{\OmegaLambda}{\ensuremath{0.692}\xspace}

\newcommand{\vdag}{(v)^\dagger}
\newcommand\aastex{AAS\TeX}
\newcommand\latex{La\TeX}

\title[O2-H0]{A gravitational-wave measurement of the Hubble constant following the second observing run of Advanced LIGO and Virgo}

\date{\today}

\begin{abstract}
This paper presents the gravitational-wave measurement of the Hubble constant $H_0$ using the detections from the first and second observing runs of the Advanced LIGO and Virgo detector network. The presence of the transient electromagnetic counterpart of the binary neutron star GW170817 led to the first standard-siren measurement of $H_0$. Here we additionally use binary black hole detections in conjunction with galaxy catalogs and report a joint measurement.
Our updated measurement is {$H_0=$ {\HubbleMeasCombinedCounterpartFLATLOG} \kmsMpc} ($68.3\%$ highest density posterior interval with a flat-in-log prior) which is {an improvement by a factor of {1.04} (about {4\%})} over the GW170817-only value of {{\HubbleMeasGWOneSevenZeroEightOneSevenCounterpartFLATLOG} \kmsMpc}.
A significant additional contribution currently comes from GW170814, a loud and well-localized detection from a part of the sky thoroughly covered by the Dark Energy Survey.
With numerous detections anticipated over the upcoming years, an exhaustive understanding of other systematic effects are also going to become increasingly important. 
These results establish the path to cosmology using gravitational-wave observations with and without transient electromagnetic counterparts.

\end{abstract}

\author{B.~P.~Abbott}
\affiliation{LIGO, California Institute of Technology, Pasadena, CA 91125, USA}
\author{R.~Abbott}
\affiliation{LIGO, California Institute of Technology, Pasadena, CA 91125, USA}
\author{T.~D.~Abbott}
\affiliation{Louisiana State University, Baton Rouge, LA 70803, USA}
\author{S.~Abraham}
\affiliation{Inter-University Centre for Astronomy and Astrophysics, Pune 411007, India}
\author{F.~Acernese}
\affiliation{Dipartimento di Farmacia, Universit\`a di Salerno, I-84084 Fisciano, Salerno, Italy}
\affiliation{INFN, Sezione di Napoli, Complesso Universitario di Monte S.Angelo, I-80126 Napoli, Italy}
\author{K.~Ackley}
\affiliation{OzGrav, School of Physics \& Astronomy, Monash University, Clayton 3800, Victoria, Australia}
\author{C.~Adams}
\affiliation{LIGO Livingston Observatory, Livingston, LA 70754, USA}
\author{R.~X.~Adhikari}
\affiliation{LIGO, California Institute of Technology, Pasadena, CA 91125, USA}
\author{V.~B.~Adya}
\affiliation{OzGrav, Australian National University, Canberra, Australian Capital Territory 0200, Australia}
\author{C.~Affeldt}
\affiliation{Max Planck Institute for Gravitational Physics (Albert Einstein Institute), D-30167 Hannover, Germany}
\affiliation{Leibniz Universit\"at Hannover, D-30167 Hannover, Germany}
% ...

\collaboration{The LIGO Scientific Collaboration and the Virgo Collaboration}

\acrodef{GW}[GW]{Gravitational wave}
\acrodef{EM}[EM]{electromagnetic}
\acrodef{CMB}[CMB]{cosmic microwave background}
\acrodef{SN}[SN]{supernovae}
\acrodef{MDC}[MDC]{mock data challenge}
\acrodef{BNS}[BNS]{binary neutron star}
\acrodef{BBH}[BBH]{binary black hole}
\acrodef{LCDM}[$\Lambda$CDM]{$\Lambda$-cold-dark-matter}
\acrodef{SNR}[SNR]{signal-to-noise ratio}
\acrodef{O2}[O2]{second observing run}
\acrodef{O3}[O3]{third observing run}
\acrodef{F2Y}[F2Y]{First Two Years}
\acrodef{GRB}[GRB]{gamma-ray burst}
\acrodef{H0}[$H_0$]{Hubble constant}
\acrodef{PSD}[PSD]{power-spectral-density}
\acrodef{SNR}[SNR]{signal-to-noise ratio}

\acrodef{DES}[DES]{Dark Energy Survey}
\acrodef{SDSS}[SDSS]{the Sloan Digital Sky Survey}
\acrodef{GLADE}[GLADE]{Galaxy List for the Advanced Detector Era}
\acrodef{GWENS}[GWENS]{Gravitational Wave Events in Sloan}

\section{Introduction}
\acp{GW} from compact binary coalescences allow for a direct measurement of the
luminosity distance to their source. This makes them standard-distance
indicators, and in conjunction with an identified host galaxy or a set of
possible host galaxies, they can be used as standard sirens to construct a
redshift-distance relationship and measure cosmological parameters like the \acl{H0}~$H_0$
\citep{Schutz:1986gp,Holz:2005df,MacLeod:2007jd,Nissanke:2009kt,Sathyaprakash:2009xt}. The \ac{GW} signature from the \ac{BNS} merger GW170817,
along with its coincident \ac{EM} transient associated with the host
galaxy NGC4993, led to a first standard-siren measurement of $H_0$
\citep{Abbott:2017xzu}. This measurement is independent of other
state-of-the-art measurements of $H_0$, and in particular, independent of the
cosmic distance ladder used to calibrate standardizable sources like Type Ia
supernovae. The importance of an independent measurement of $H_0$ is worth
highlighting. With the Planck 2018 data release \citep{Aghanim:2018eyx}, and the recalibration of supernovae using Large Magellanic Cloud Cepheids
\citep{Riess:2019cxk}, the tension between early universe measurements of $H_0$ from Planck and local measurements from the SH0ES project has risen to the $4.4$-$\sigma$ level. Independent measurements using cosmological Baryon Acoustic Oscillations to calibrate Type Ia supernovae via the inverse distance ladder \citep{Macaulay:2018fxi} and gravitational lensing of quasars in the nearby universe \citep[H0LiCOW Collaboration,][]{Birrer:2018vtm} favor to some degree the early-universe Planck and the local SH0ES measurements respectively. A complementary measurement of $H_0$ from the multi-messenger \ac{GW} astronomy
sector
would help clarify whether the current tension is a statistical anomaly or evidence for new physics beyond the $\Lambda$CDM model of cosmology.\footnote{Cosmological parameters can potentially be inferred from \ac{GW} observations alone by estimating the redshift using the known physics of neutron stars \citep{Messenger:2011gi} or their astrophysical mass distribution \citep{Finn:1994cg,Taylor:2012db}; however these methods are not expected to find an application in context of the current generation of advanced ground-based detectors. {Prospects of \ac{GW} cosmology using robust astrophysical features of black hole mass distributions have recently been explored in \cite{Farr:2019twy}.}}

The \ac{GW} standard-siren measurement in \cite{Abbott:2017xzu} is
broadly consistent with other measurements. By
combining information from multiple detections, one can improve the accuracy reaching about one percent with $\mathcal{O}(100)$
detections in the coming years
\citep{Dalal:2006qt,Nissanke:2013fka,Chen:2017rfc,Feeney:2018mkj,Vitale:2018wlg,Mortlock:2018azx}.

An unambiguous identification of the host galaxy is unlikely for all \ac{BNS} detections; only a crude estimate of the sky position may be
available. Moreover there
are sources such as \ac{BBH} mergers with no expected \ac{EM} counterparts.
Even in the absence of an \ac{EM} counterpart, the method outlined in
\cite{Schutz:1986gp} can be used: with a set of potential host galaxies
identified in a galaxy catalog for each detection, one can build up information
by a process of statistical cross-correlation. The method was demonstrated on a
set of simulations in \cite{DelPozzo:2011yh}, where a $5\%$ estimate on $H_0$
was obtained from $\mathcal{O}(100)$ detections in an idealized situation of
nearby events and complete galaxy catalogs; similar results with projections
for third-generation detectors have been obtained in \cite{Nair:2018ign}. It
has further been shown in \cite{Chen:2017rfc} that the main benefit of the
galaxy-catalog method would be for the case of multiple well-localized sources. 

An understanding of \ac{GW} selection effects
\citep{Abbott:2017xzu,Chen:2017rfc,Mandel:2018mve,Vitale:2020aaz} and features of galaxy
catalogs, such as their incompleteness and measurement uncertainties, is
necessary for an accurate measurement of $H_0$.  
Prescriptions to handle incomplete galaxy catalogs have been outlined in \cite{Chen:2017rfc}, \cite{Fishbach:2018gjp}, and \cite{Gray:2019ksv}, and an extensive
study of selection
effects including galaxy catalog completeness has been performed in \cite{Gray:2019ksv}. The simulations in \cite{Gray:2019ksv} demonstrate that the current method is able to handle galaxy catalog completeness down to $\sim 25\%$ for $\mathcal{O}(100)$ detections in the nearby universe ({luminosity distance} $d_\mathrm{L} \lesssim 115\, \Mpc$) with idealized galaxy catalogs, without introducing additional systematic features. 
The galaxy-catalog method {was} used in \cite{Fishbach:2018gjp} to infer $H_0$ from GW170817 without its optical counterpart. An estimate of $H_0$ from GW170814 and the photometric
redshift catalog from the \ac{DES} Year 3 data {was also} obtained in
\cite{Soares-Santos:2019irc}.

In this paper we report the first joint \ac{GW} estimate of $H_0$
from detections during O1 and O2, the first and second observing runs of the
Advanced LIGO and Virgo detector network.
{For our final result, along with the \ac{BNS} GW170817, we {choose the six} O1 and O2 \ac{BBH} detections {which satisfy our selection criterion of a network signal-to-noise ratio $\text{SNR}>12$ in at least one search pipeline for \ac{GW} detections}. The detections for which we have significant galaxy catalog support are GW150914,
GW151226, GW170608, and GW170814. For these events, we expect the inferred $H_0$ posteriors to be driven by the additional \ac{EM} information. 
The {two remaining \acp{BBH} which satisfy the selection criterion, GW170104 and GW170809, need to be retained in the analysis for consistency with the assumed population model. These events} for which a significant fraction of potential host galaxies are missing in the associated galaxy catalog, {can} also {potentially} contribute to the $H_0$ measurement, since there is information about cosmology present in the observed source distribution \citep{Taylor:2011fs,Taylor:2012db,Farr:2019twy}.
The distribution of the observed source parameters, including the observed luminosity distance distribution, is driven by $H_0$ in addition to the intrinsic astrophysical source distributions.
With a known underlying astrophysical distribution of sources, $H_0$
would be measureable solely from the observed distance distribution of \ac{BBH} detections \citep{Chen:2014yla}.}
This {latter contribution} is not expected to provide significant information at this stage
given how uncertain the inferred {astrophysical distribution parameters such as the power law index of the binary mass distribution, or the index characterizing the evolution of the binary merger rate with redshift} are, even when $H_0$ is held fixed \citep{LIGOScientific:2018jsj}. 
In an ideal situation, one {should} jointly estimate {these} astrophysical population parameters along with $H_0$.
We choose to fix the astrophysical population to a fiducial distribution instead, and perform our
analysis with different choices for the mass distribution and binary merger
rate evolution with redshift in order to quantify possible systematic effects resulting from this assumption. 
We set aside a more thorough treatment involving the marginalization over the unknown astrophysical distribution for future work.

The main result of our analysis---a posterior distribution on $H_0$---is 
dominated by the contribution from GW170817 with its
optical counterpart, {with a modest improvement from the inclusion of the
O1 and O2 \acp{BBH}.} These results, possibly refined and marginalized over
the aforementioned assumptions, can be used as a prior for future
\ac{GW} estimates of $H_0$. The analysis performed in this paper
thus serves as a precursor of future analyses for the third and
subsequent observing runs of the Advanced detector network. 

The rest of this paper is arranged as follows. We describe our method in Section~\ref{sec:method}. We summarize the \ac{GW} detections we use in our analysis and the corresponding \ac{EM}
data in Section~\ref{sec:data}. Our main results are presented in
Section~\ref{sec:results}, with a more detailed discussion and a study of possible systematic effects in Section~\ref{sec:discussion}. We
conclude in Section~\ref{sec:conclusion} and highlight some future directions and prospects.  

Throughout this paper we assume a $\Lambda$CDM cosmology and use
the best-fit Planck 2015 values of \mbox{$\Omega_m=\OmegaMatter$},
\mbox{$\Omega_\Lambda=\OmegaLambda$}, respectively for the fractional matter
and dark energy densities in the present epoch \citep{Ade:2015xua}.
Although these parameters enter the redshift-distance relationship central to the method for Bayesian inference of \ac{H0},
we have verified that our results are robust with regards to a variation of their values within the current measurement uncertainties.

\section{Method}
\label{sec:method}
We follow and apply the Bayesian analysis described in  \cite{Gray:2019ksv} to compute the posterior probability density on $H_0$, given the set $\{D_\text{GW}\}$ of $N_\text{det}$ detections 
and the associated \ac{GW} data $\{x_\text{GW}\}$:
\begin{equation}
p(H_0|\{x_{\text{GW}}\},\{D_{\text{GW}}\}) \propto p(H_0)p(N_{\text{det}}|H_0)\prod_i^{N_{\text{det}}} p({x_{\text{GW}}}_i|{D_{\text{GW}}}_i,H_0)\,.
\end{equation}
Here, $D_{\text{GW}i}$ indicates that the event $i$ was detected as a GW, $p(H_0)$ is the prior on $H_0$, and 
the term $p(N_{\text{det}}|H_0)$ is the likelihood of detecting $N_{\text{det}}$ events {over the time of observation with the associated detector configuration} for the given value of $H_0$. 
{The number of detected events is some fraction of the total number of sources $N_s$, and this fraction depends on $H_0$; thus $N_\text{det} = f_\text{det}(H_0)\,N_s$.
If the intrinsic astrophysical merger rate,
$R$, which $N_s$ is proportional to},\footnote{The astrophysical merger rate is defined as $R\equiv{\partial{N_{s}}}/{(\partial V\, \partial T)}$, where $N_s$ is the number of sources as in the text, $V$ the comoving sensitive volume, and $T$ the time of observation or survey.} is marginalized over with a
prior $p(R) \propto R^{-1}$, then $p(N_\text{det}|H_0)=\int p(N_\text{det}|H_0,R)\,p(R)\,dR$
{is independent of $f_\text{det}(H_0)$, and thus} 
loses its dependence on $H_0$ \citep[{see, e.g.,}][]{Fishbach:2018edt}.
For simplicity, we make this approximation throughout our analysis.
The final term factorises into the individual likelihoods for each detection. In the following, we write out the expressions for a single \ac{GW} event $i$, omitting the subscript $i$ for brevity of notation,
\begin{equation}
\begin{aligned}\label{Eq:factorlikelihood}
p(x_{\text{GW}}|D_{\text{GW}},H_0) &= \dfrac{p(D_{\text{GW}}|x_{\text{GW}},H_0)p(x_{\text{GW}}|H_0)}{p(D_{\text{GW}}|H_0)}\,.
\end{aligned} 
\end{equation}
The denominator, $p(D_{\text{GW}}|H_0)$, is evaluated as an integral over all possible $x_{\text{GW}}$ \citep{Abbott:2017xzu,Chen:2017rfc,Mandel:2018mve}:
\begin{align}
\label{Eq:detectionefficiency}
p(D_\text{GW}|H_0)=\int p(D_\text{GW}|x_\text{GW},H_0)\,p(x_\text{GW}|H_0)\,dx_\text{GW}\,,
\end{align}
where $p(D_{\text{GW}}|x_\text{GW},H_0)=1$ in the case where the \ac{SNR} of $x_{\text{GW}}$ passes some detection threshold, and 0 in the case where it does not.

\subsection{The electromagnetic counterpart case}
\label{sec:counterpart}
In the presence of an \ac{EM} counterpart, there is additional information in the \ac{EM} data, {$x_\text{EM}$,} which appears as an \ac{EM} likelihood term. {Together with this, there is an underlying assumption
that there has been an \ac{EM} detection; we denote this assumption as $D_\text{EM}$.} 
Thus, for a single event with an \ac{EM} counterpart,
\begin{equation}
\begin{aligned}
p(x_{\text{GW}},x_{\text{EM}}|D_{\text{GW}},D_{\text{EM}},H_0) &= \dfrac{p(x_{\text{GW}}|H_0 ) p(x_{\text{EM}}|H_0 )}{p(D_{\text{EM}}|D_{\text{GW}},H_0 ) p(D_{\text{GW}}|H_0 )}\,.
\end{aligned} 
\end{equation}
We assume that the detectability of an \ac{EM} counterpart is dependent on luminosity distance (as opposed to redshift) because it is flux-limited.  As \ac{GW} detectability is also a function of luminosity distance, we expect $p(D_{\text{EM}}|D_{\text{GW}},H_0)$ to be a constant that does not depend on $H_0$. This leads to
\begin{equation}
\begin{aligned}
p(x_{\text{GW}},x_{\text{EM}}|D_{\text{GW}},D_{\text{EM}},H_0) &\approx \dfrac{p(x_{\text{GW}}|H_0 ) p(x_{\text{EM}}|H_0 )}{p(D_{\text{GW}}|H_0 )}.
\end{aligned} 
\end{equation}

\subsection{The galaxy-catalog case}
\label{sec:statistical}
In the absence of an \ac{EM} counterpart, the analogous data comes from galaxy catalogs which provide a set of galaxies and their associated sky locations, redshifts, and apparent magnitudes.  As we are in the regime where the detectability of \ac{GW} sources extends beyond the distance to which current catalogs are complete, the possibility that the \ac{GW} host galaxy is not contained in the catalog, because it is too faint, has to be taken into account. This is done by marginalizing over the cases where the host is in the catalog (denoted $G$), and where it is not (denoted $\bar{G}$):
\begin{equation} \label{Eq:Likelihood}
\begin{aligned}
p(x_{\text{GW}}|D_{\text{GW}},H_0)
&= \, p(x_{\text{GW}}|G,D_{\text{GW}},H_0) p(G|D_{\text{GW}}, H_0)\\
&+p(x_{\text{GW}}|\bar{G},D_{\text{GW}},H_0) p(\bar{G}|D_{\text{GW}}, H_0)\,.
\end{aligned} 
\end{equation}
{In the case where the host galaxy is in the catalog, the EM data enters in the form of information available in the galaxy catalog, as also in \cite{Fishbach:2018gjp} and \cite{Soares-Santos:2019irc}. This assumption is included in our conditioning on $G$. The EM information is used to modify our priors on galaxy redshift, sky location, and (apparent) magnitude, which are common among all GW events using the same catalog. This differs from the counterpart case where the \ac{EM} data enters the expression as a likelihood term, $x_{\text{EM}}$, a transient which is informative for only one GW event. In the case where the host galaxy is not in the catalog, the complementary condition $\bar{G}$ implies that we include the information about the limitations of the \ac{EM} survey. Following \cite{Gray:2019ksv}, we model the galaxy catalog as having an apparent magnitude threshold, $m_\text{th}$, since galaxy catalogs are flux-limited. This magnitude threshold, alongside the intrinsic (absolute) brightness of a galaxy and its luminosity distance, determines the probability that the galaxy is inside or outside the galaxy catalog.

The formalism which we adopt from \cite{Gray:2019ksv} is closely analogous to the method described in \cite{Chen:2017rfc} and \cite{Fishbach:2018gjp}. The in-catalog and out-of-catalog terms in Eq.~\eqref{Eq:Likelihood} above correspond to the two terms in Eq. (3) of \cite{Fishbach:2018gjp}. However the methods start formally differing in the way that the (in)completeness of the galaxy catalog is treated. 
In the current formalism from \cite{Gray:2019ksv}, the (in)completeness of the galaxy catalog follows naturally from the parameters of the underlying EM survey(s). The assumption that a galaxy catalog can be modelled by a single $m_\text{th}$ in each direction of the sky is an idealization. However this can in future be extended to include non-trivial selection functions modeling the optical sensitivity of the telescope instead of a sharp cutoff at $m_\text{th}$, and to composite catalogs coming from multiple surveys.}

The quantities appearing on the right in Eq.~\eqref{Eq:Likelihood} can be written out explicitly as follows. 
\begin{equation} \label{Eq:p(x|G,D,H0)}
\begin{aligned}
p(& x_{\text{GW}}|G, D_{\text{GW}}, H_0)  \\
&= \dfrac{\sum^{N_\text{gal}}_{j=1} \int p(x_{\text{GW}}|z_j,\Omega_j,H_0)p(s|M(z_j,m_j,H_0)) p(z_j) dz_j}
{\sum^{N_\text{gal}}_{j=1} \int p(D_{\text{GW}}|z_j,\Omega_j,H_0)p(s|M(z_j,m_j,H_0)) p(z_j) dz_j}\,.
\end{aligned}
\end{equation}
Here, $N_\text{gal}$ is the total number of galaxies in the galaxy catalog, $\Omega_j$ and $m_j$ are respectively the sky coordinates and apparent magnitude for galaxy $j$, and $p(z_j)$ is a distribution representing the redshift of galaxy $j$. {This quantity, $p(z_j)$, which enters as a prior for our analysis, is the posterior distribution on the galaxy redshift obtained following analysis of EM data; details of the profiles chosen for $p(z_j)$ are deferred until Section \ref{sec:z-umcertainties}.} The quantity $M(z_j,m_j,H_0)$ is the absolute magnitude (for the given $H_0$), and $p(s|M(z_j,m_j,H_0))$ is the probability of a galaxy with these parameters to host a \ac{GW} source during the observation time, relative to other galaxies. Formally, $s$ is the statement that a GW has been {\em sourced} or {\em emitted} (as opposed to being {\em detected}); the previous expressions are all implicitly conditioned on the assumption of $s$. In writing $p(s|M)$, we make the approximation that the probability of a galaxy hosting a source depends only on the intrinsic luminosity of the galaxy, and not on its other parameters or on the properties of the \ac{GW} source.
In essence, this term allows for weighting  galaxies by their luminosities $L(M_j(H_0))$ as
\begin{equation} \label{Eq:luminosityweighting}
\begin{aligned}
p(s|M(z_j,m_j,H_0)) &\propto 
\begin{cases}
\text{constant} & \text{if unweighted}\\
L(M_j(H_0)) & \text{if luminosity-weighted.}
\end{cases}
\end{aligned}
\end{equation}
The likelihood when the host galaxy is not in the catalog, $p(x_{\text{GW}}|\bar G,D_{\text{GW}},H_0)$, is a ratio of marginalized integrals:
\begin{widetext}
\begin{equation} \label{Eq:p(x|Gbar,D,H0)}
\begin{aligned}
&p(x_{\text{GW}}|\bar G,D_{\text{GW}},H_0)
= \dfrac{\iiint^\infty_{z(m_{\text{th}},M,H_0)}
p(x_{\text{GW}}|z,\Omega,H_0) p(z)p(\Omega)p(M|H_0)p(s|M) dz d\Omega dM}{\iiint^\infty_{z(m_{\text{th}},M,H_0)} 
p(D_{\text{GW}}|z,\Omega,H_0) p(z)p(\Omega)p(M|H_0)p(s|M) dz d\Omega dM}\,.
\end{aligned}
\end{equation}
Here the fact that the terms are conditioned on $\bar{G}$ is incorporated into the redshift limits as a function of the apparent magnitude threshold $m_\text{th}$ of the galaxy catalog.
Finally, the prior probabilities that a given \ac{GW} detection has or does not have support in the galaxy catalog are respectively
\begin{equation} \label{Eq:p(G|D,H0)}
\begin{aligned}
p(G|D_{\text{GW}}, H_0)
&= \dfrac{\iiint^{z(m_{\text{th}},M,H_0)}_0 
p(D_{\text{GW}}|z,\Omega, H_0) p(z)p(\Omega)p(M|H_0 )p(s|M)dz  d\Omega  dM }
{\iiint_0^\infty p(D_{\text{GW}}|z,\Omega, H_0) p(z)p(\Omega)p(M|H_0 )p(s|M) dz d\Omega dM},\quad \text{and} \
p(\bar{G}|D_{\text{GW}}, H_0) = 1 - p(G|D_{\text{GW}}, H_0).
\end{aligned}
\end{equation}
\end{widetext}
In Eqs.~\eqref{Eq:p(x|Gbar,D,H0)} and~\eqref{Eq:p(G|D,H0)}, $p(z)$ is the prior on the redshift
of host galaxies of GW events,
taken to be of the form
\begin{equation}
p(z) \propto \frac{1}{1+z}\frac{dV_\text{c}(z)}{dz} R(z)\,.
\end{equation}
Here $V_\text{c}(z)$ is the comoving volume as a function of redshift
and the factor $(1+z)^{-1}$ converts the merger rate {from the source frame to the detector frame}. The merger rate density may in general be a function of redshift; however we set $R(z) = \text{constant}$ throughout (other than in Section~\ref{sec:discussion}, where we consider an alternative redshift-dependent rate model). The prior on the \ac{GW} sky location $p(\Omega)$ is taken to be uniform across the sky.
The term $p(M|H_0)$ is the prior on absolute magnitudes for all the galaxies in the universe (not just those inside the galaxy catalog), which we set to follow the Schechter luminosity function:
\begin{equation}
\label{eq:schechter}
p(M|H_0)\propto 10^{-0.4(\alpha+1)(M-M^*(H_0))}\exp\left[{-10^{-0.4(M-M^*(H_0))}}\right].
\end{equation}
Following \cite{Gehrels:2015uga}, we use $B$-band luminosity function parameters $\alpha=-1.07$ for the slope of the Schechter function and {$M^*(H_0)=-19.7+5\log_{10}h$} for its characteristic absolute magnitude
(with $h \equiv H_0 / 100 \, \kmsMpc$) throughout, {unless otherwise specified}.\footnote{The absolute magnitude is related to the intrinsic luminosity of a galaxy by the relation, $M-M^*\equiv-2.5\log_{10}(L/L^*)$. The parameter $M^*$ of the Schechter function itself depends on $H_0$, which we take into account.}
For the upper limits of integration over $M$, we choose the magnitude of the dimmest galaxies to be {$-12.2 + 5\log_{10}h$}. The integrals are not sensitive to the choice of their lower limits, \ie the magnitudes of the brightest galaxies. More complex models for $p(M|H_0)$ can be used; in fact, we expect the luminosity distribution of galaxies to also evolve with redshift \citep{1989ApJ...337L..65C}, as well as to depend on galaxy type and color \citep{2002MNRAS.333..133M}. While the consideration of such dependence is beyond the scope of the current work, we refer the reader to \cite{Gray:2019ksv} for a brief discussion on the misspecification of the luminosity function parameters.

Further details and complete derivations for the framework described above are discussed in \cite{Gray:2019ksv}.

{
\subsection{Posterior samples re-weighting}
The \ac{GW} data $x_{\text{GW}}$ has undergone parameter estimation, providing a set of posterior samples on source parameters, including the luminosity distance $d_\mathrm{L}$, sky-coordinates $\Omega$, and the observed component masses in the detector frame $m_{1,2}^{\text{det}}$. We need the likelihood of the \ac{GW} data $p(x_{\text{GW}}|H_0)$, which requires removing the priors used for the parameter estimation.  In particular, this means that the $d_\mathrm{L}^2$ prior must be removed before the priors on $z$ and $H_0$ can be applied.

Additionally, a more subtle effect comes into play regarding source- and detector-frame masses. The treatment of \ac{GW} selection effects requires a marginalisation over the source-frame mass distribution, while the data $x_{\text{GW}}$ contains information about the detector-frame masses.  In order to be self-consistent, the analysis must be performed using the same prior assumptions on both the individual event data, and the normalizing $p(D_{\text{GW}}|H_0)$ term.  This requires removing the detector-frame mass prior and re-weighting the posterior samples with the source-frame mass prior.

The source-frame mass distribution is $p(m_1, m_2 |\, \mu)$ where $\mu$ denotes the hyper-parameters describing the astrophysical model (concrete details of the assumed model are discussed later in Section~\ref{sec:results}). To re-weight the likelihood we use the measured detector-frame masses and luminosity distances to compute the corresponding source-frame masses as a function of $H_0$,
\begin{equation}
m_{1,2} = \frac{m_{1,2}^{\text{det}}}{1+z\,(d_\mathrm{L},H_0)}\,.
\end{equation}
We marginalize out the source-frame mass dependence, using the prior $p(m_1, m_2 |\, \mu)$.  The \ac{GW} likelihood is a function of $d_\mathrm{L}$ and $\Omega$ as before, but now is also a function of $H_0$. This effect will be most pronounced for high distance (redshift) events, where the conversion between detector-frame and source-frame masses is most noticeable.  Additionally, posterior samples from events with particularly high or low detector-frame masses may become inconsistent with the source-frame mass prior for certain values of $H_0$, leading to additional constraining power (see Section \ref{sec:pop} for details).}

\section{Data}
\label{sec:data}
\subsection{Gravitational-wave data}
\label{sec:detections}
The \ac{GW} searches performed during the first and the second observation runs of Advanced LIGO and Virgo have led to the identification of ten \ac{BBH} and one \ac{BNS} mergers \citep{LIGOScientific:2018mvr}. The \ac{BNS} event GW170817, well-localized and at a nearby distance of \bnsdistance \Mpc, helped discover the electromagnetic transient from the merger, and was subsequently associated with host galaxy NGC4993. The \acp{BBH} span a large range of distances from \closest \Mpc to \farthest \Mpc and are distributed over the sky with $90\%$ credible regions as low as \mosttight \sqdeg to as high as \leasttight \sqdeg. A summary of the relevant parameters of all the \ac{GW} detections are given in Table~\ref{table:gwdata}. 

{For the main results presented in Section~\ref{sec:results} of this work, we choose the events which meet the selection criterion of network SNR $> 12$ in at least one of the two pipelines for modeled searches, namely PyCBC and GstLAL. The sensitivity of the results to this choice is explored in Section~\ref{sec:discussion} by reducing the network SNR threshold to $11$. In each case, this SNR threshold is also used in the treatment of GW selection effects as described in Eq.~\eqref{Eq:detectionefficiency} above.} In practice a detection is claimed not solely on the basis of the \ac{SNR}, but additionally by applying data quality vetoes in order to remove noise transients, and eventually constructing a ranking statistic such as an inverse false alarm rate or a likelihood-ratio \citep{LIGOScientific:2018mvr}. {A careful treatment should use a threshold on a ranking statistic rather than the \ac{SNR} as the selection criterion. However a distinction between the two does not cause an appreciable difference if the considered detections are significantly louder than transient noise artifacts \citep[see, \eg Appendix A.1 of][]{LIGOScientific:2018jsj}. In order to keep the computation of the GW selection effects tractable, one can thus place a more stringent threshold on the SNR and select a subset of loud events from the detected population, which is what we do for our analysis. We note that placing an SNR threshold lower than or close to that set by the detection piplines would again be problematic without modifications to the current method of accounting for GW selection effects.}

\begin{table*}
\centering
{\renewcommand{\arraystretch}{1.5}
\renewcommand{\tabcolsep}{0.2cm}
\begin{tabular}{cccccccccc}
\hline
\hline
Event & {SNR} & $\Delta\Omega/$\sqdeg & $d_\mathrm{L}/$\Mpc & $z_\text{event}$ & $V/\Mpc^3$ & Galaxy catalog & Number of galaxies & {$m_{\text{th}}$} & $p(G|z_{\text{event}},D_{\text{GW}})$ \\
\hline
GW150914 & {24.4} & 182 & $440^{+150}_{-170}$ & $0.09^{+0.03}_{-0.03}$ & $3.5\times 10^6$ & GLADE & 3910 & {17.92} & $0.42$\\
GW151012 & {10.0} & 1523 & $1080^{+550}_{-490}$ & $0.21^{+0.09}_{-0.09}$ & $5.8\times 10^8$ & GLADE & 78195 & {17.97} & $0.01$\\
GW151226 & {13.1} & 1033 & $450^{+180}_{-190}$ & $0.09^{+0.04}_{-0.04}$ & $2.4\times 10^7$ & GLADE & 27677 & {17.93} & $0.41$\\
GW170104 & {13.0} & 921 & $990^{+440}_{-430}$ & $0.20^{+0.08}_{-0.08}$ & $2.4\times 10^8$ & GLADE & 42221 & {17.76} & $0.01$\\
GW170608 & {15.4} & 392 & $320^{+120}_{-110}$ & $0.07^{+0.02}_{-0.02}$ & $3.4\times 10^6$ & GLADE & 6267 & {17.84} & $0.60$\\
GW170729 & {10.8} & 1041 & $2840^{+1400}_{-1360}$ & $0.49^{+0.19}_{-0.21}$ & $8.7\times 10^9$ & GLADE & 77727 & {17.82} & $<0.01$\\
GW170809 & {12.4} & 308 & $1030^{+320}_{-390}$ & $0.20^{+0.05}_{-0.07}$ & $9.1\times 10^7$ & GLADE & 18749 & {17.62} & $<0.01$\\
GW170814 & {16.3} & 87 & $600^{+150}_{-220}$ & $0.12^{+0.03}_{-0.04}$ & $4.0\times 10^6$ & DES-Y1 & {31554} & {23.84} & {$>0.99$}\\
GW170817 & {33.0} & 16 & $40^{+7}_{-15}$ & $0.01^{+0.00}_{-0.00}$ & $227$ & -- & -- & {--} & --\\
GW170818 & {11.3} & 39 & $1060^{+420}_{-380}$ & $0.21^{+0.07}_{-0.07}$ & $1.5\times 10^7$ & {GLADE} & {1059} & {17.51} & {$<0.01$}\\
GW170823 & {11.5} & 1666 & $1940^{+970}_{-900}$ & $0.35^{+0.15}_{-0.15}$ & $3.5\times 10^9$ & GLADE & 117680 & {17.98} & $<0.01$\\
\hline
\end{tabular}}
\caption{Relevant parameters of the O1 and O2 detections: {network signal-to-noise ratio (SNR) for the search pipeline (PyCBC/GstLAL) in which the signal is the loudest,} $90\%$ sky localization region $\Delta\Omega$ (\sqdeg), luminosity distance $d_\mathrm{L}$ (\Mpc, median with $90\%$ credible intervals), and estimated redshift $z_\text{event}$ (median with $90\%$ range assuming Planck 2015 cosmology) from \cite{LIGOScientific:2018mvr}. In the remaining columns we report the corresponding $90\%$ 3D localization comoving volumes, the number of galaxies within each volume for public catalogs which we find to be the most complete, {and the apparent magnitude threshold, $m_{\text{th}}$, of the galaxy catalog associated with the corresponding sky region (as described in Section~\ref{sec:pcat}).} The final column gives the probability that the host galaxy is inside the galaxy catalog for each event, $p(G|z_{\text{event}},D_{\text{GW}})$, also evaluated at the median redshift for each event.}
\label{table:gwdata}
\end{table*}

\subsection{Galaxy Catalogs}
\label{sec:catalogs}

The analysis with \acp{BBH} is performed in conjunction with appropriate galaxy catalogs. 
We use the GLADE catalog \citep{Dalya:2018cnd} as a default, due to its depth and coverage over an extensive region of the sky (see Section~\ref{sec:GLADE}). 
For \ac{GW} observations that are particularly well-localized, certain galaxy catalogs show a clear improvement in completeness over GLADE within the relevant localization volume of the event.
In particular, we use the DES Year 1 (Y1A1 GOLD or simply Y1) catalog \citep{Drlica-Wagner:2017tkk,Abbott:2018jhe} for the analysis of GW170814 {(see Section~\ref{sec:DESY1})}.
{GW170818 lies within the footprint of \ac{SDSS}. While not used in this work, galaxy catalogs based on Data Release 14 \citep{Abolfathi:2017vfu} or Data Release 16 \citep{Ahumada:2019vht} of \ac{SDSS} \citep[including curated versions such as the GWENS catalog,][]{GWENS}, could be used to improve the current analysis for events which fall in the SDSS footprint.}

In Table~\ref{table:gwdata} we summarize the galaxy catalogs that we use for our analysis for each of the detections, along with the number of galaxies in the $90\%$ error volume calculated from 3D skymaps constructed from posterior samples associated with the data release of \cite{LIGOScientific:2018mvr},\footnote{Available at: \url{https://www.gw-openscience.org/GWTC-1}} {and estimates of the probability that the host galaxy is in the catalog, evaluated at the median redshift for each detection} assuming a Planck 2015 cosmology.

In the following, we describe in more detail the galaxy catalogs that we use, quantify the probability that the host galaxy for each event is in the galaxy catalog that is used for its analysis and discuss the assessment of the completeness over the relevant localization volume for the best localized events. Finally, we quantify the uncertainties associated with the photometric measurement of redshifts in some of these catalogs.

\subsubsection{GLADE}
\label{sec:GLADE}
We use the \ac{GLADE} version {2.4} galaxy catalog \citep{Dalya:2018cnd},\footnote{GLADE is publicly available at: \url{http://glade.elte.hu}} to construct the observed redshift distributions for the majority of the detected \acp{BBH}. The GLADE catalog has an all sky coverage \citep[Fig.~1 of][]{Dalya:2018cnd} since it is constructed from the GWGC \citep{White:2011qf}, 2MPZ \citep{Bilicki:2013sza}, 2MASS XSC \citep{Skrutskie:2006wh}, HyperLEDA \citep{Makarov:2014txa} and SDSS-DR12Q \citep{Paris:2016xdm} catalogs. The GLADE catalog is complete (in $B$-band luminosity) out to 37 Mpc and has an estimated completeness of 50\% out to 91 Mpc \citep[Fig.~2 of][]{Dalya:2018cnd}. At low redshifts ($\lesssim 0.05$), we expect to be dominated by the peculiar velocity field. GLADE reports peculiar-velocity-corrected redshifts following the reconstruction of \cite{Carrick:2015xza}. GLADE provides apparent magnitudes in the $B$-band, which we can use directly (\ie without any photometric transformations) for luminosity weighting of the galaxies.

\subsubsection{DES Year 1}
\label{sec:DESY1}
The Dark Energy Survey (DES) is a five year survey {mapping} $\approx$ 300 million galaxies in five filters ({\em grizY}) over $5000~\sqdeg$. It is worth noting that the GW170814 sky localization is fully enclosed within the footprint of the \ac{DES} \citep{Drlica-Wagner:2017tkk,Abbott:2018jhe} Year 3 (Y3) ``gold'' catalog. An estimate of $H_0$ from the GW170814 distance and the Y3 catalog of the \ac{DES} has been carried out \citep{Soares-Santos:2019irc}. In this work, we use the publicly available \ac{DES}-Y1 catalog \citep{Abbott:2018jhe},\footnote{\ac{DES}-Y1 is available at: \url{https://des.ncsa.illinois.edu/releases/y1a1}} to compute the $H_0$ posterior for GW170814. {Approximately} 87\% of the probability region for the GW170814 sky localization is enclosed within the \ac{DES}-Y1 catalog. Analysis with a different catalog provides a parallel measurement of $H_0$ with GW170814, and (given the catalog differences) can potentially be indicative of systematic effects in the catalogs, such as the treatment of redshift uncertainties (provided that a similar set of galaxies are present in both catalogs). 

{We select the objects in the DES-Y1 catalog that are classified as high-confidence galaxies using the default  classification scheme, ``MODEST\_CLASS'' \citep{Drlica-Wagner:2017tkk,Sevilla-Noarbe:2018ktd}. We use the photometric redshifts which are derived using the Bayesian Photometric Redshift (BPZ) template fitting method \citep{Hoyle:2017mee}}. We use the median redshifts provided in the catalog and discard (around {5\%) galaxies with redshift errors larger than twice their corresponding quoted median redshift value. Such a choice is not expected to bias our result since the discarded galaxies are highly uninformative}. 

{We convert from the DES {\em grizY} magnitudes to the SDSS {\em ugriz} system using the photometric transformations provided in the DES-Y1 paper \citep{Drlica-Wagner:2017tkk}, which requires discarding a further $\sim5$\% of galaxies with inadequate color information. This transformation enables us to apply K corrections to obtain source-frame luminosities (see Sec. \ref{sec:Kcorr} for details).  We use the SDSS $g$-band magnitudes, as these are closest to $B$-band, and update the Schechter parameters of our analysis to have $\alpha=-0.89$ and  $M^*(H_0)=-19.39+5\log_{10}h$ based on \cite{Blanton_2003}.}

\subsection{Probability that the host galaxy is in the catalog}
\label{sec:pcat}
\begin{figure*}
\centering
\includegraphics[width=\linewidth]{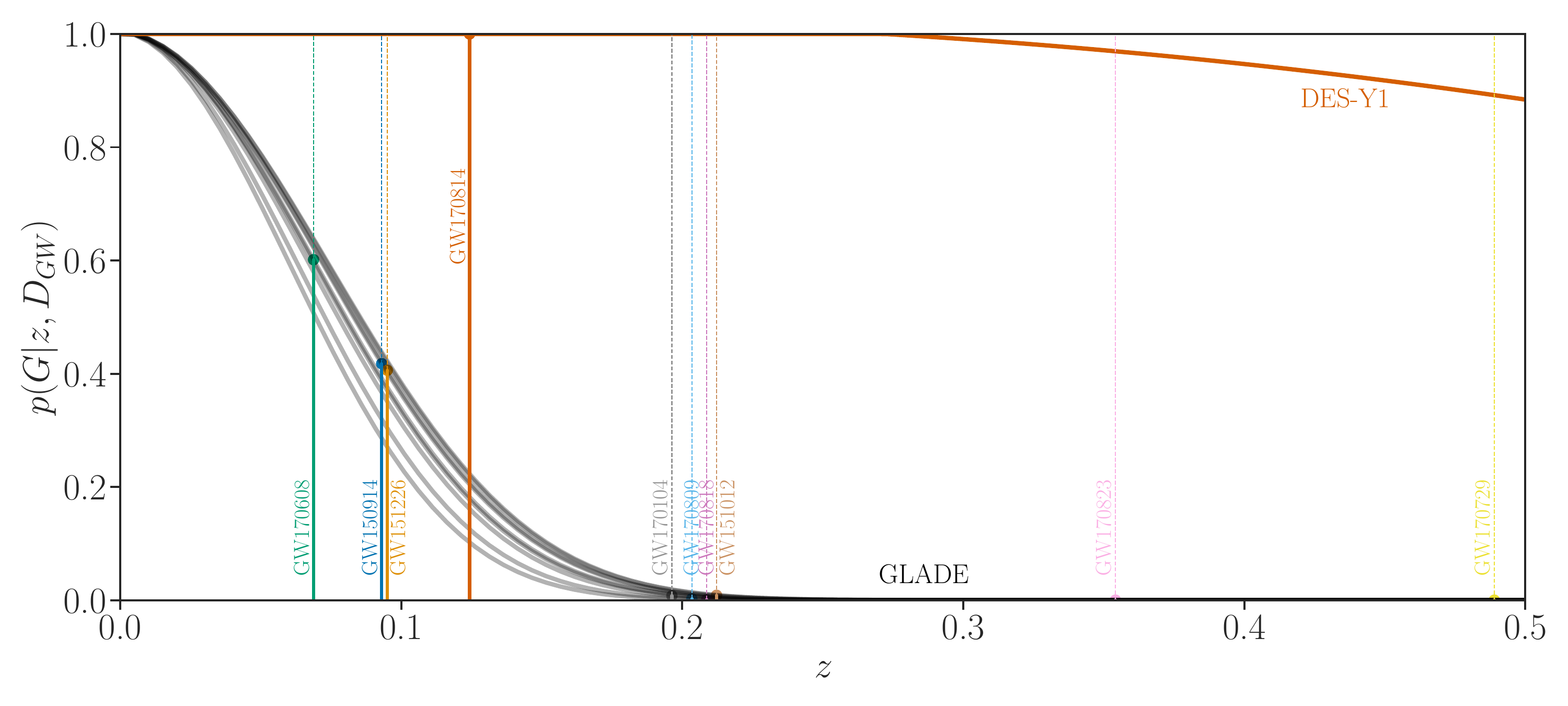}
\caption{The probability that the host galaxy is inside the galaxy catalog, shown for \ac{GLADE} ({gray curves}) and \ac{DES}-Y1 (orange curve), as a function of redshift.  For \ac{GLADE} this quantity is calculated {for each individual event, using the completeness estimated within each event's sky localization}.  For \ac{DES}-Y1 the curve is only valid in the patch of sky covering GW170814. Each curve is independent of the value of \ac{H0}. The vertical lines show the median redshift (assuming a Planck 2015 cosmology) for each event as in Table~\ref{table:gwdata}.  These lines are thick and solid up to the intercept with the galaxy catalog they are used with, and thin and dashed above.  
}
\label{Fig:catalog:all}
\end{figure*}

In this work, we assume that we can characterize the completeness of a galaxy catalog using an apparent magnitude threshold (limiting magnitude) $m_{\text{th}}$. We estimate $m_{\text{th}}$ by calculating the median value from the apparent magnitude distribution of all the galaxies within the {sky localization of each event.  For GLADE, this choice allows us to account for some of the larger changes in completeness across the sky, which come from it being a composite catalog, comprised of many surveys of differing depths.} 
Galaxy catalogs are directional, and a more sophisticated analysis would involve calculating the limiting magnitude for a given line of sight. Obtaining the $H_0$ posterior distribution would thus require a joint estimate of $m_{\text{th}}$ along the lines of sights within an event's sky localization. We leave this for future work. That the completeness of a galaxy catalog is modelled by a set of limiting magnitude thresholds, can by itself be a non-trivial assumption, especially for photometric catalogs, since galaxies may be missing for various reasons other than them being too faint. This will also need to be revisited in the future in a catalog-specific manner.

For now, we use the $m_{\text{th}}$ estimated as described above, and show in Fig.\ref{Fig:catalog:all} the probability of a host galaxy being inside the catalog $p(G|z,D_{\text{GW}})$ as a function of redshift $z$, for each of the galaxy catalogs under consideration. {For \ac{GLADE} this quantity is calculated for each event using the $m_{\text{th}}$ calculated for each event's sky localization.} For \ac{DES}-Y1, the curve is for the patch of sky covering GW170814. {These probabilities are calculated using the expressions in Eq.~\eqref{Eq:p(G|D,H0)}, but as a function of $z$, and are therefore independent of the choice of \ac{H0}}. We additionally show as the vertical lines in Fig.~\ref{Fig:catalog:all} the median redshift for each event $z_\text{event}$ (calculated assuming a Planck 2015 cosmology). In the final two columns of Table~\ref{table:gwdata}, we report the $m_\text{th}$ {of the relevant catalog within the sky localization of each event, and the probability of the host galaxy being in the catalog at the median redshift of each event.}

\subsection{Detailed analysis of DES-Y1}
\label{sec:detailedanalysis}
\begin{figure}
\centering
\includegraphics[width=\linewidth]{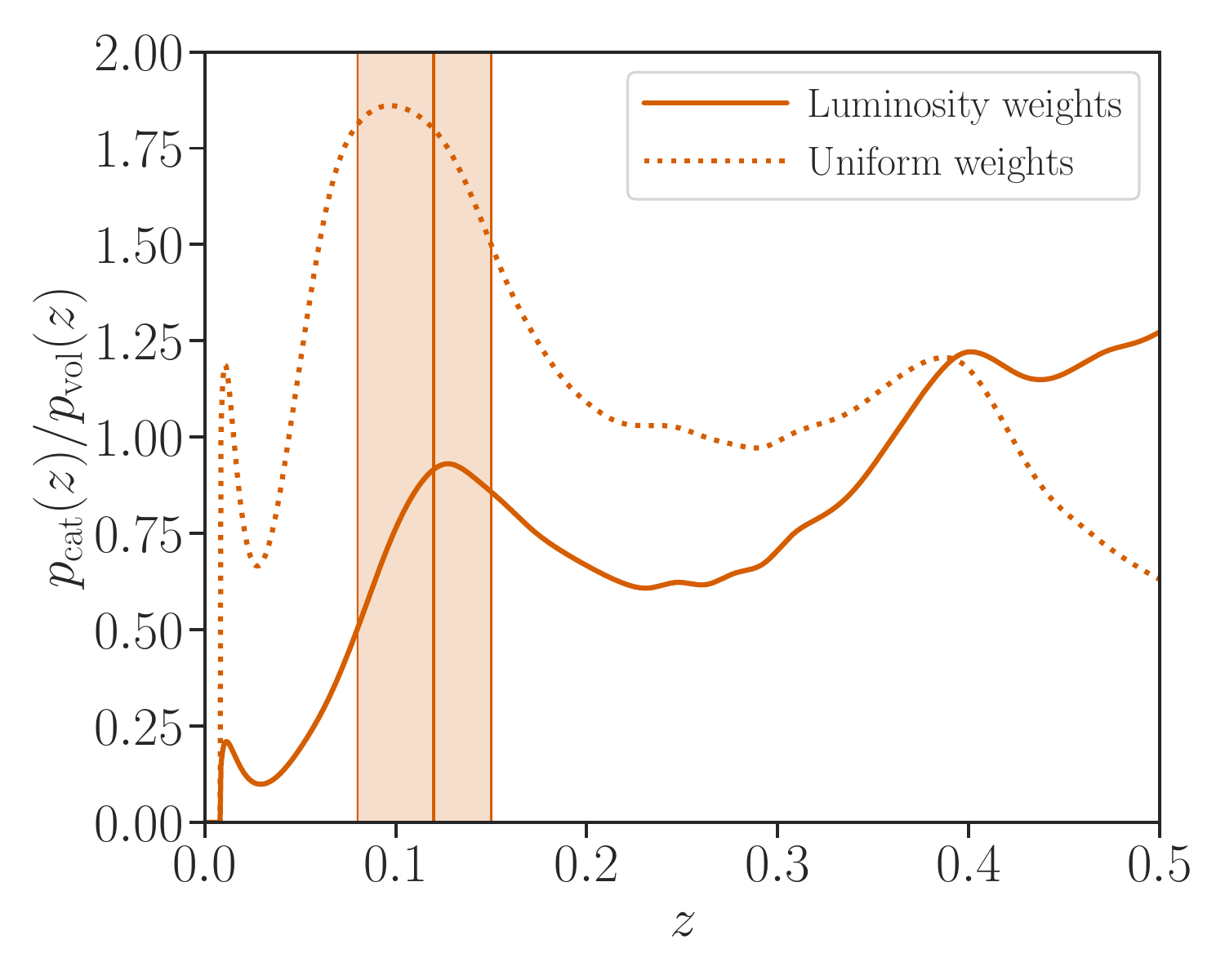}
\caption{
Probability distribution for the redshifts of potential host galaxies $p_\text{cat}(z)$, {with redshift uncertainties taken into account,} divided out by a uniform in comoving volume distribution $p_\text{vol}(z)$ of galaxies. When computing $p_\text{cat}(z)$ we include all galaxies brighter than {$0.05 L_g^*$} within the corresponding event's 99\% sky localization region and weight each galaxy by weights proportional to their $g$-band luminosity (solid lines) as well as with uniform weights ({dotted} lines). We show these distributions for the DES-Y1 galaxies within the GW170814 sky localization region. 
We also show the 90\% median estimated redshift range for GW170814 (calculated assuming a Planck 2015 cosmology) for reference.}
\label{Fig:catalog:gwensdes}
\end{figure}

The high completeness fraction of DES-Y1 within the GW170814 {sky localization} is apparent from Fig.~\ref{Fig:catalog:all}.
The catalog is expected to be more complete than GLADE since it has a limiting magnitude of approximately {23.8 for DES-Y1. We analyze the \ac{EM} information coming from this catalog in greater detail.}
It is helpful to have {an} assessment of the contribution from potential host galaxies as a function of redshift for these events. {In order to quantify this contribution, we} perform a treatment analogous to \cite{Fishbach:2018gjp} and compute the ratio $p_\text{cat}(z)/p_\text{vol}(z)$ between the probability distribution for the redshifts of potential host galaxies $p_\text{cat}(z)$ and of a uniform in comoving volume distribution of galaxies $p_\text{vol}(z)$. When computing $p_\text{cat}(z)$ we include all galaxies brighter than {$0.05 L_g^*$} within the corresponding event's 99\% sky localization region defined as,
\begin{equation}
p_\text{cat}(z) \equiv \int p(x_\text{GW}|\Omega)\,p_0(z,\Omega)\,d\Omega\,,
\end{equation}
where $p(x_\text{GW}|\Omega)$ is the \ac{GW} likelihood as a function of the sky position $\Omega$ (this effectively weights each galaxy with the 2D skymap probability), and $p_0(z,\Omega)$ represents the galaxy catalog contribution, obtained from the distribution of galaxies in the catalog, {marginalized over their redshift uncertainties also obtained from the catalog, and} weighted by their probability of hosting a GW source (assuming a Planck 2015 cosmology for the required magnitude conversion). We consider weights for each galaxy proportional to their $g$-band luminosity as well as uniform weights to explore the effects due to this choice. 

In Fig.~\ref{Fig:catalog:gwensdes} we show the distributions $p_\text{cat}(z)/p_\text{vol}(z)$ for the DES-Y1 galaxies within the GW170814 sky localization region, for the redshift range $0 < z < 0.5$. 
{The unweighted curve traces the over/under-density of galaxies, and then falls off at larger redshift due to incompleteness in the catalog. The luminosity-weighted redshift distribution is driven partially by the over-density of galaxies at $z \approx 0.4$, and partially by bright high-redshift galaxies.}
The host galaxies for GW170814 are more likely to be located near {the} higher galaxy density regions in the DES-Y1 catalog -- these features in the redshift prior are expected to drive the inferred $H_0$ posteriors for the corresponding events. {Features} we see in the DES-Y1 catalog are not as pronounced as the overdensity in the {DES-Y3} data seen in \cite{Soares-Santos:2019irc}. {While the {DES-Y3} survey is deeper, and may reveal finer features, a part of the above difference is likely also driven by the difference in the photometric redshift estimation algorithms, namely, {template fitting methods such as BPZ \citep{Hoyle:2017mee} and machine learning based methods such as the \texttt{ANNz2} algorithm \citep{Sadeh:2015lsa}, with the latter used for GW170814 in \citep{Soares-Santos:2019irc}}. Only the former of the two has been used for the DES-Y1 catalog and a combination of both for the DES-Y3 catalog.} The different selection criteria for choosing galaxies from the two catalogs, such as the stringent redshift cut placed in \cite{Soares-Santos:2019irc} versus a more relaxed redshift prior used in this work, is another potential source of difference between the corresponding redshift distributions.

\subsubsection{Redshift uncertainties}
\label{sec:z-umcertainties}
An important source of measurement uncertainty with galaxy catalogs is {in the galaxy redshifts, which are often photometric estimates} due to lack of spectroscopic measurements out to large redshifts. In our formalism, the uncertainty in redshift is reflected in the term $p(z_j)$ in Eq.~\eqref{Eq:p(x|G,D,H0)}. {We model the individual galaxy redshift probability distributions, $p(z_j)$, as a Gaussian distribution with mean set to the quoted median photometric redshift $z_\text{photo}$, and with standard deviation of $\sigma_{z_\text{photo}}$ (both obtained from the galaxy catalog data)}. Photometric redshift estimates are often not approximated by Gaussian distributions, and we make this assumption only due to limited information present in some of the public galaxy catalogs which we use. A rigorous quantification of the effect of this assumption is beyond the scope of this work.
{In practice, we implement the redshift uncertainty by the process of a Monte Carlo integration: the integral over $z_j$ in Eq.~\eqref{Eq:p(x|G,D,H0)} becomes an additional sum over $N_\text{photo}$ random samples. This number varies between the different galaxies, and is always high enough to ensure that the final event likelihood is independent of the set of random draws from each galaxy.}

{
\subsubsection{Source-frame luminosities}
\label{sec:Kcorr}
For galaxy catalogs which are complete to high redshifts, it is important to note that the observed apparent magnitudes are redshifted from their source frame. Galaxies do not emit uniformly in all wavelengths, and galaxy surveys are only sensitive in certain bands. In order to compare the properties of galaxies at different redshifts, e.g.~to apply luminosity weighting, or to apply a luminosity cut to the galaxy sample, the $K$ correction, described in \cite{Oke1968,Hogg:2002yh}, needs to be applied. In particular, the $K$ correction term is redshift dependent, and so neglecting it could lead to systematic bias in the estimation of $H_0$.

For DES, we compute the $K$ corrections using the calculator described in \cite{Kcorcalc2010,Kcorcalc2012}. These $K$ corrections are valid out to redshifts of 0.5, and so we apply a hard redshift cut. In order to remain self-consistent, the method is adjusted to include this redshift cut, such that the probability of the host galaxy being in the catalog above $z=0.5$ is zero, and below the cut it is determined according to the apparent magnitude threshold, as before.

In the case where galaxies have high redshifts and large redshift uncertainties, this uncertainty must be accounted for in the calculation of the $K$ corrections. Using only the mean redshift to calculate the $K$ correction would lead to unrealistically bright or faint luminosities at the tail ends of the redshift distribution. In our implementation, the $K$ corrections are calculated in-situ for each redshift Monte Carlo random draw described in \ref{sec:z-umcertainties}, automatically taking the galaxy redshift uncertainties into account.
}

\section{Results}
\label{sec:results}
We apply the method described above to obtain a measurement of the Hubble constant using \ac{GW} standard sirens only. 
We carry out our analysis with a prior on $H_0$ uniform in the interval of $[\HubblePriorMin,\,\HubblePriorMax]~\kmsMpc$; we report our final results also using a flat-in-log prior $p(H_0)\propto H_0^{-1}$ in the same interval for ease of comparison with previous studies. We use the marginalized distance likelihood and skymaps constructed from the posterior samples of \cite{LIGOScientific:2018mvr}.\footnote{For computational convenience, we separately construct a marginalized distance likelihood and a two-dimensional skymap; this approximation will be revisited in the future.} For the \acp{BBH}, we choose all galaxies in the $99.9\%$ sky region of the corresponding catalog {and we further weight the galaxies in proportion to their $B(g)$-band luminosities.} In order to calculate the term $p(D_{\text{GW}}|H_0)$ in the denominator, we use a Monte Carlo integration, sampling parameters which affect an event's detectability (masses, sky location, inclination angle, and polarisation) from chosen priors. We choose a power-law mass distribution for \acp{BBH} with $p(m_{1})\propto m_{1}^{-\alpha}$ and $m_2$ uniform in its range with {$5M_\odot < m_2 < m_1 < 100M_\odot$} in the source frame, and a distribution of merger rates that does not evolve with redshift; for the power-law index $\alpha$, we choose $\alpha=1.6$ \citep[which is supported by Model B of][]{LIGOScientific:2018jsj}. For \acp{BNS}, we use a Gaussian mass distribution with a mean of $1.35 M_\odot$ and a standard deviation of $0.15 M_\odot$ \citep{Kiziltan:2013oja}.\footnote{While this does not significantly affect our current results, we will need to revisit the BNS mass distribution in light of GW190425 \citep{Abbott:2020uma}, which is potentially a \ac{BNS} merger with heavier components.} The remaining \ac{GW} parameters are marginalized over their natural distributions: uniform in the sky, uniform on the sphere for orientation, uniform in polarization. We use the time-averaged \acl{PSD} of detector noise for the corresponding observation run from \cite{Aasi:2013wya}, and for the selection criterion, we use a {network \ac{SNR} threshold of $12$ in at least one search pipeline. The O1 and O2 \acp{BBH} which pass this criterion are GW150914, GW151226, GW170104, GW170608, GW170809, and GW170814 (see Table~\ref{table:gwdata}).}

\begin{figure*}
\centering
\includegraphics[width=\linewidth]{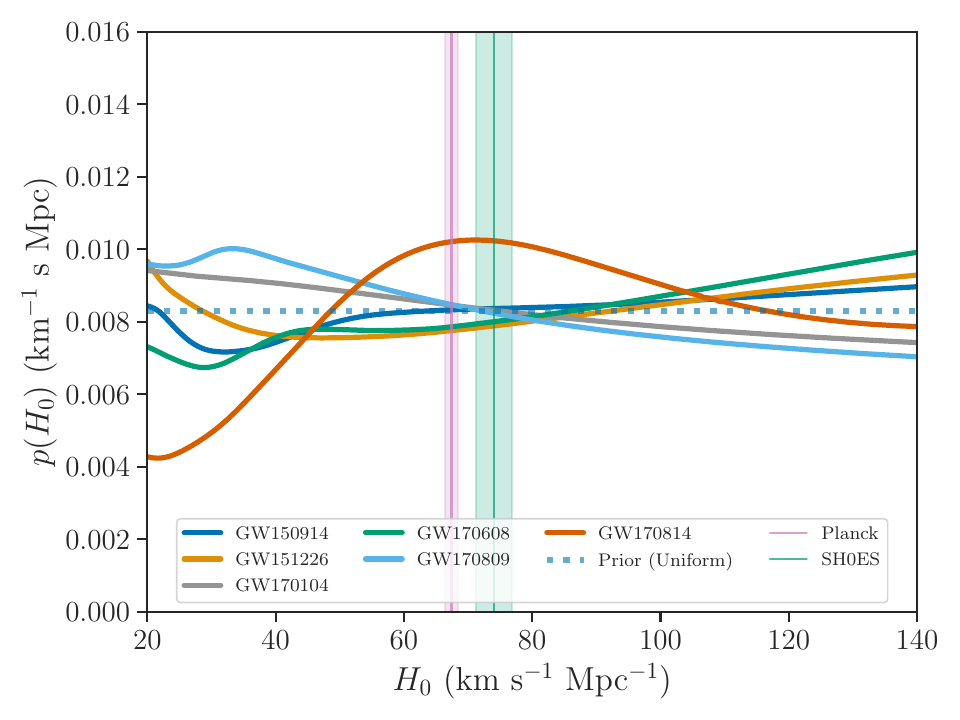}
\caption{Individual estimates of $H_0$ from the {six \acl{BBH} detections which satisfy the selection criterion of network $\text{SNR} > 12$ in at least one search pipeline}. These results assume a $m^{-1.6}$
power-law distribution on masses and a non-evolving rate model.
All results assume a prior on $H_0$ uniform in the interval [\HubblePriorMin,\,\HubblePriorMax]\,\kmsMpc (dotted blue).
We also show the estimates of $H_0$ from CMB \citep[Planck:][]{Aghanim:2018eyx} and supernova observations \citep[SH0ES:][]{Riess:2019cxk}.
}
\label{Fig:posterior:jointbbh} \end{figure*}

Our result for {these} O1 and O2 {\acp{BBH}} is shown in Fig.~\ref{Fig:posterior:jointbbh}. The detections for which there is considerable support from the galaxies present in the catalog show features of the galaxy catalog in their $H_0$ posterior distribution. The GW170814 estimate is qualitatively similar to the result in \citet{Soares-Santos:2019irc} with analogous peaks in the posterior distribution. The differences in peak locations can be attributed to a difference in the redshift distribution for the DES-Y3 catalog used in \cite{Soares-Santos:2019irc} versus that for the public DES-Y1 catalog used in this work. For the detections for which the galaxy catalogs are relatively empty, we see the features of the assumptions on mass distribution and redshift evolution of binary merger rate that have entered our analysis. The more distant events such as {GW170809} lead to $H_0$ estimates pushed to the lower end of the prior. This is due to the following reason. In a universe where host galaxies are distributed uniformly in comoving volume, for lower values of $H_0$, the expected distribution of detected GW events favors relatively higher luminosity distances. Thus events at high luminosity distances have more support for smaller values of $H_0$, while relatively nearby events, namely GW150914, GW151226, GW170608 and GW170814, correspondingly have lower support at smaller values of $H_0$. 
The information present in the observed luminosity distance distribution would thereby potentially contribute to the $H_0$ measurement, independent of, or even in absence of information in galaxy catalogs, if the underlying distributions of source parameters were known.
{In the absence of knowledge of the astrophysical distribution of \ac{BBH} source parameters, a thorough treatment would involve a marginalization over all possible mass distributions and rate models. The following section discusses the systematic differences that the assumptions on the assumed population model could lead to.}

\begin{figure*}
\centering
\includegraphics[width=\textwidth]{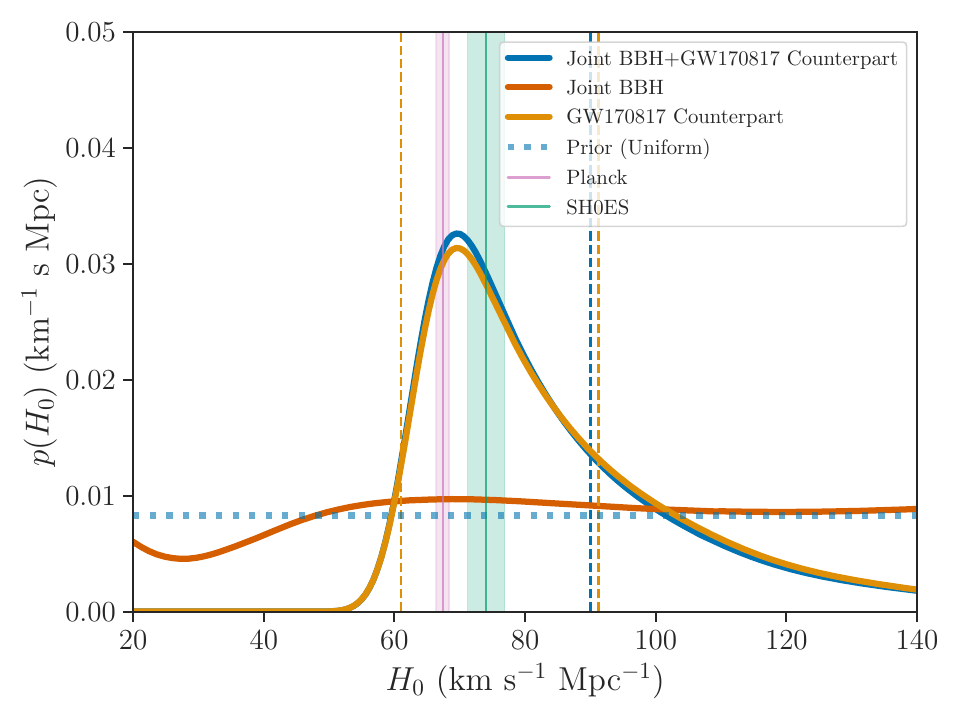}
\caption{The gravitational-wave measurement of $H_0$ (dark blue) from the detections in the first two observing runs of Advanced LIGO and Virgo. The GW170817 estimate (orange) comes from the identification of its host galaxy NGC4993 \citep{Abbott:2017xzu}. The additional contribution comes from \aclp{BBH} in association with appropriate galaxy catalogs; for GW170814 we use the DES-Y1 galaxy catalog, {while for the remaining {five} \acp{BBH}, {GW150914, GW151226, GW170104, GW170608, and GW170809,} we use the GLADE catalog}. The $68\%$ maximum a-posteriori intervals are indicated with the vertical dashed lines. All results assume a prior on $H_0$ uniform in the interval [\HubblePriorMin,\,\HubblePriorMax]\,\kmsMpc (dotted blue). We also show the estimates of $H_0$ from CMB \citep[Planck:][]{Aghanim:2018eyx} and supernova observations \citep[SH0ES:][]{Riess:2019cxk}.}
\label{Fig:posterior:jointcounter}
\end{figure*}

For our final result we combine the contribution of the \acp{BBH} with the result from GW170817 obtained using the low spin prior samples from \cite{LIGOScientific:2018mvr} and an estimated Hubble velocity of $v_H\equiv cz= 3017 \pm 166\, \text{km} \text{s}^{-1}$ (where $c$ is the speed of light) for NGC4993 from \cite{Abbott:2017xzu}.
Our final combined result is shown in Fig.~\ref{Fig:posterior:jointcounter}, with the posterior distribution plotted assuming a uniform $H_0$ prior: we obtain {$H_0=$ {\HubbleMeasCombinedCounterpart} \kmsMpc} ($68.3\%$ highest density posterior interval).
To compare with values in the literature, we also use a flat-in-log prior, $p(H_0)\propto H_0^{-1}$, and calculate 
{$H_0=$ {\HubbleMeasCombinedCounterpartFLATLOG} \kmsMpc}, which corresponds to an improvement by a factor of {1.04} (about {4\%}) over the GW170817-only value of {{\HubbleMeasGWOneSevenZeroEightOneSevenCounterpartFLATLOG} \kmsMpc}. {We also quote the median and symmetric 90\% credible interval for this measurement, which is {\HubbleMeasCombinedCounterpartFLATLOGMEDIAN} \kmsMpc.}

\section{Systematic effects}
\label{sec:discussion}
In this section, we repeat the {previous} analysis with {a few} alternative {assumptions for {the \ac{GW} selection criterion,} the \ac{GW} population model, and source host properties}. {Subsequently, we discuss other sources of systematic effects, a detailed study of which is beyond the scope of this work.}

\subsection{Selection criterion}
\label{sec:snr}
\begin{figure}
\includegraphics[width=0.49\textwidth]{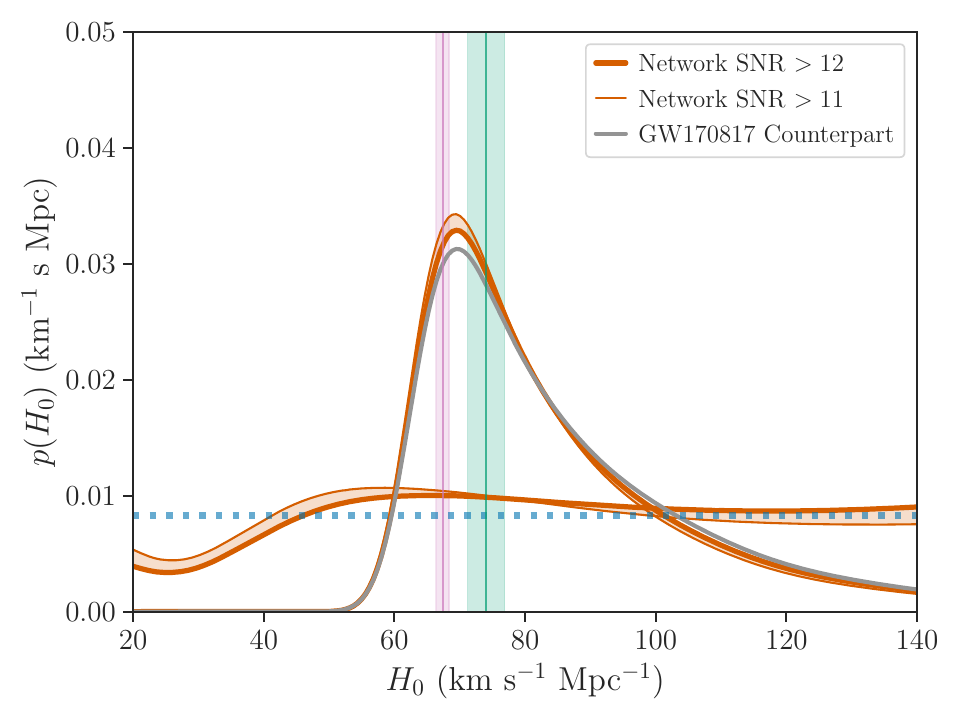}
\caption{Sensitivity of the results to the selection criterion on the signal-to-noise ratio (SNR). The final joint result as well as the contribution from all the \acp{BBH} which satisfy the selection criterion are shown for a threshold network SNR of $12$ (thick orange) and $11$ (thin orange); the variation is shown as a shaded band. The GW170817 counterpart result (gray) is added to guide the eye. Six \acp{BBH} (GW150914, GW151226, GW170104, GW170608, GW170809, and GW170814) pass the selection criterion with $\text{SNR}>12$. Two additional \acp{BBH} (GW170818 and GW170823) are included with $\text{SNR}>11$. These results assume the default $m^{-1.6}$ power-law distribution on masses and a non-evolving rate model.}
\label{Fig:posterior:snr}
\end{figure}

{We first look into the sensitivity of our results to the \ac{GW} selection criterion. We reduce the threshold on the network SNR from $12$ chosen in the previous section to $11$ in least one search pipeline. The computation of the \ac{GW} selection effects is adjusted accordingly. Fig.~\ref{Fig:posterior:snr} shows the results with the two sets of assumptions. Reducing the SNR threshold to 11 introduces two additional events in our analysis, namely GW170818 and GW170823, neither of which have a significant in-catalog contribution. Differences are expected due to the fact that additional low-SNR events are introduced, and also because the individual likelihoods change slightly with a different SNR threshold used in the GW selection term. In the regime of a large number of events, these two effects are expected to cancel, provided that the additional low-SNR events do not have significant in-catalog support.
For our result, this difference is not significant; however the small variation is a reminder that we are still in the regime of a low number of events.}

\subsection{Population model}
\label{sec:pop}
\begin{figure*}
\centering
\includegraphics[width=0.49\textwidth]{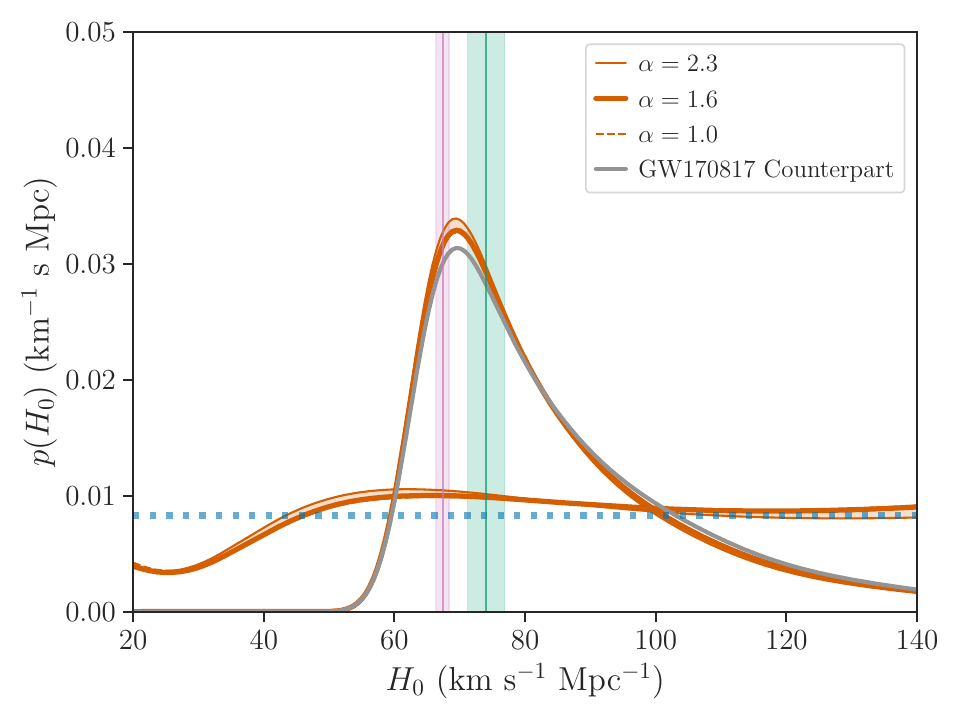}
\includegraphics[width=0.49\textwidth]{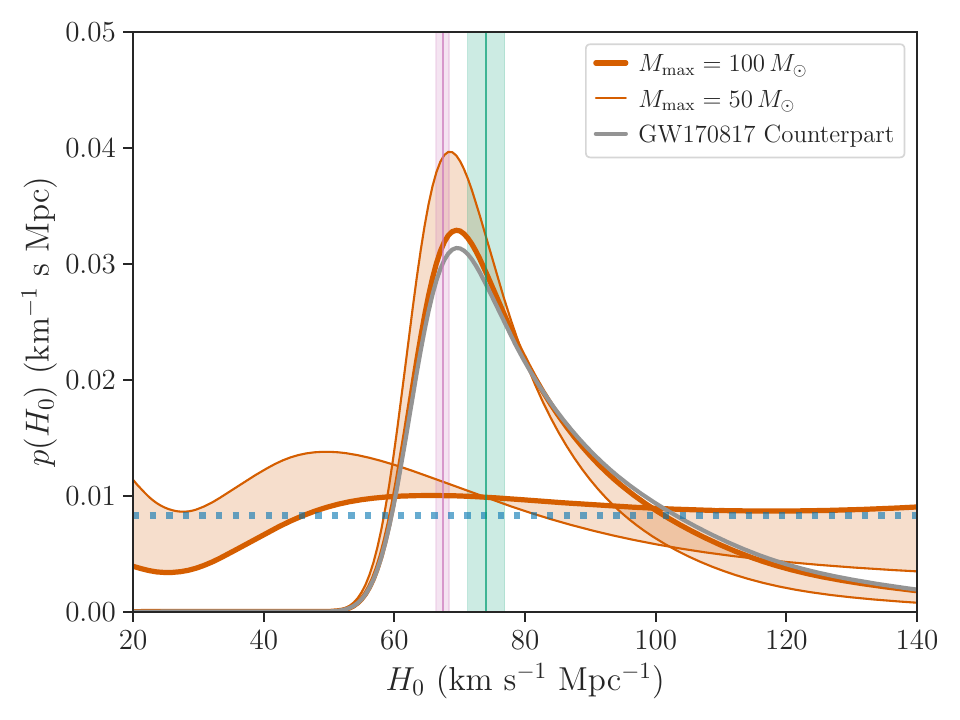}
\caption{Sensitivity of the results to the assumed mass distribution model. {\em Left panel:} Variation of the results with three different choices of the power-law index for the mass distribution, $\alpha=1.6$ (thick solid), $\alpha=2.3$ (thin solid) and $\alpha=1$ (thin dashed) assuming a constant intrinsic astrophysical merger rate, $R(z)=\text{constant}$ and $M_\text{max} = 100\,M_{\odot}$. {{\em Right panel:} Variation of the results with two different choices for the allowed black hole maximum mass, $M_\text{max} = 100\,M_{\odot}$ (thick solid), and $M_\text{max} = 50\,M_{\odot}$ (thin solid), both assuming $R(z) =$ constant and $\alpha=1.6$.}}
\label{Fig:posterior:mass}
\end{figure*}

{Going back to our default assumption of a network SNR threshold of $12$, we} test the sensitivity to our assumptions regarding the population model, \ie the mass distribution and the distribution of binary merger rate with redshift.
In addition to the power-law mass distribution with $\alpha=1.6$ \citep[median inferred value using Model B of][]{LIGOScientific:2018jsj}, we choose a shallower flat-in-log mass distribution with $\alpha=1$, and a steeper distribution with $\alpha=2.3$ \citep[both within the support of the inferred range in][]{LIGOScientific:2018jsj}. {Our results are shown in the left panel of Fig.~\ref{Fig:posterior:mass}, and they demonstrate that the systematic differences due to the choice of power-law slope $\alpha$ are insignificant.

{As a test case,} we vary the upper cut-off for the mass distribution, $M_\text{max}$. For our default analysis, $M_\text{max}$ was chosen to be $100\,M_\odot$, consistent with all {the considered} \acp{BBH} for all values of $H_0$ within the prior range. Reducing this cut-off to {a slightly restrictive} $M_\text{max} = 50\,M_\odot$ {\citep[e.g., motivated by the pair instability supernova process,][]{Fowler:1964zz},} we see a significant difference (right panel of Fig.~\ref{Fig:posterior:mass}). {Lowering $M_\text{max}$ corresponds to a closer GW detection horizon. This systematically leads each event to prefer slightly lower values of $H_0$ than in the main result, for the reasons outlined in Section~\ref{sec:results} -- namely the relationship between the predicted event distribution (from our GW selection effects) and the real detected event distribution.}}

\begin{figure}
\includegraphics[width=0.49\textwidth]{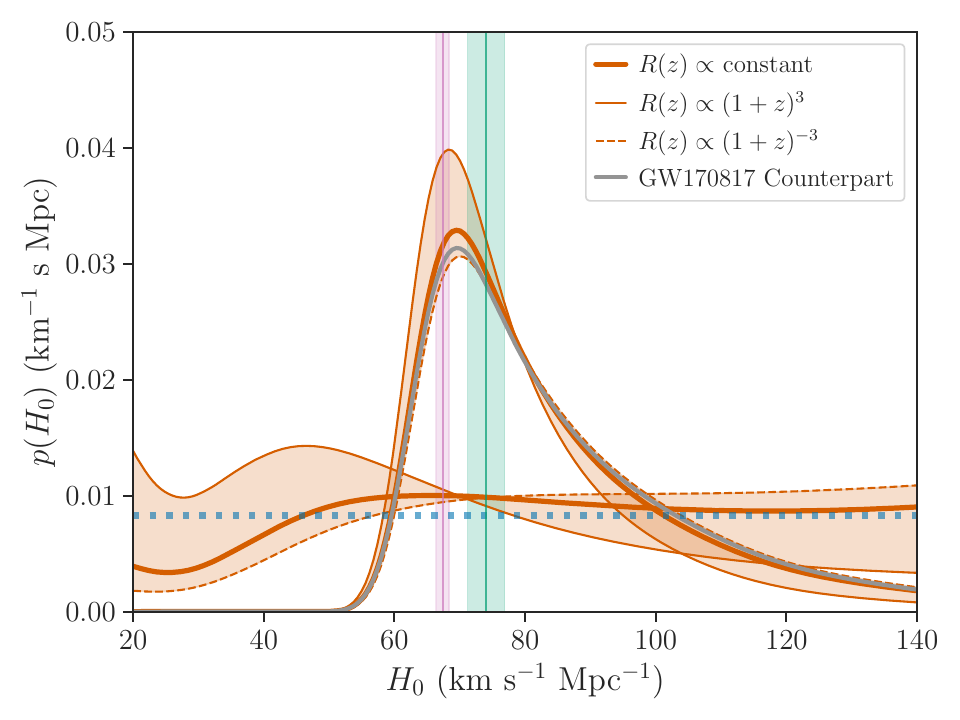}
\caption{Variation of the results with two different choices for the rate evolution, $R(z)=\text{constant}$ (thick solid), and $R(z)\propto(1+z)^3$ (thin solid) {and $R(z)\propto(1+z)^{-3}$ (thin dashed)} for $\alpha=1.6$ and $M_{max} = 100M_{\odot}$.  }
\label{Fig:posterior:rate}
\end{figure}

{Next, we} relax our assumption on the evolution of rate of binary mergers with redshift. A constant merger rate density, $R(z)=\text{constant}$, implicit in the previous treatment, assumed that the merger rate traces the comoving volume. 
In addition, we repeat our analysis using a merger rate $R(z)\propto (1+z)^3$, which traces the star formation rate at low redshifts ($z < 2.5$) \citep{10.1093/mnras/242.3.318}, {as well as a merger rate $R(z)\propto (1+z)^{-3}$ which could arise if typical delay times are very long, as may be expected from the chemically homogeneous evolution formation channel \citep{Mandel:2015qlu}}. 
These relaxed assumptions thus cover a large fraction of physically viable and inferred population models \citep{LIGOScientific:2018jsj}.
We show our results for the different assumed redshift evolution models in Fig.~\ref{Fig:posterior:rate}. {The model in which the merger rate traces star formation shows a significant difference, with a tendency to prefer lower values of $H_0$, compared to the other two models which are quite similar. This is the only systematic effect that leads to a significant difference in the results at this time.}

\subsection{Luminosity weighting}

\begin{figure}
\centering
\includegraphics[width=0.49\textwidth]{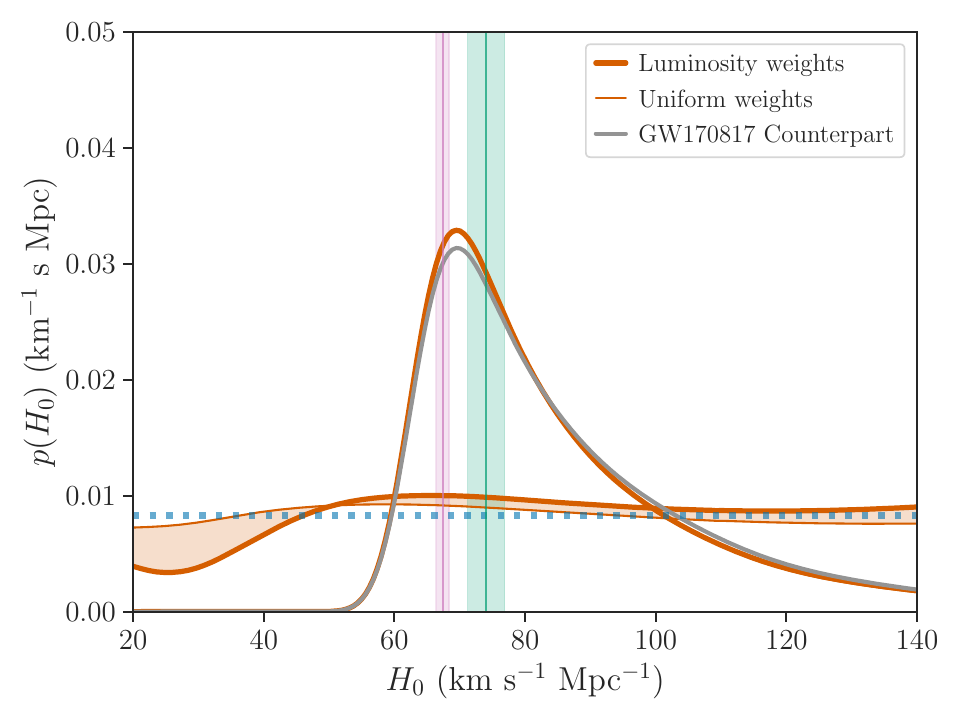}
\caption{Sensitivity of the results to luminosity weighting. We show how the results vary when we weight the galaxies in the catalog by their B(g)-band luminosity (thick solid) as well as with constant (uniform) weights (thin solid), both assuming a power-law index for the mass distribution, $\alpha=1.6$ and constant intrinsic astrophysical merger rate, $R(z)={\rm constant}$.}
\label{Fig:posterior:lumweights}
\end{figure}

The results in the previous section assumed a weighting of galaxies by their luminosities in the B-band, 
which are indicative of galaxies' star formation rates. 
In order to quantify the difference likely to be caused by alternate ways of weighting the galaxies, we repeat our analysis with no luminosity weighting. These results are shown in Fig.~\ref{Fig:posterior:lumweights}. 

{With uniform luminosity weights, we obtain a result on a joint binary black hole estimate which is close to flat (thin orange line in Fig.~\ref{Fig:posterior:lumweights}).} This can be understood as follows: 1) The out-of-catalog terms in Eq.~\eqref{Eq:Likelihood} take into account the lack of galaxies beyond the apparent magnitude threshold $m_\text{th}$ of the catalog in a uniform way. {When galaxies are unweighted, the probability of the host galaxy being inside the catalog is reduced compared to the luminosity-weighted case. More weight is given to the uniform out-of-catalog contribution, and the events become less informative.} 2) {The catalogs used in this analysis contain high numbers of low-luminosity galaxies.}  {The contribution from these more evenly distributed dim galaxies, in the unweighted case, again reduces the informativeness of each event and flattens the final result.} This is also in agreement with our expectations from \cite{Fishbach:2018gjp} and \cite{Gray:2019ksv}, where weighting by luminosities enhance the features in the posterior distribution coming from the galaxy catalog.

\subsection{Waveform models}

The posterior samples of \cite{LIGOScientific:2018mvr} used for the results in this paper have been obtained combining the results of gravitational waveform models which incorporate spin and precession effects to different extents \citep{Husa:2015iqa,Khan:2015jqa,Hannam:2013oca,Pan:2013rra,Taracchini:2013rva,Babak:2016tgq}. These models are restricted to quasi-circular orbits (i.e., they do not include orbital eccentricity) and neglect higher-order harmonics. Systematic differences in GW parameter estimation results with the employed waveform models constitute only a small fraction of the total uncertainty budget \citep[see, e.g.,][]{TheLIGOScientific:2016pea,Abbott:2016wiq}, and given the large statistical uncertainties, the ignored effects in waveform modeling are not expected to cause a difference to the current measurement of $H_0$. However cumulative systematic effects arising from limitations of waveform models will become increasingly important as the statistical uncertainties become smaller and, in particular, features that can lead to biases in the GW estimation of distance will need to be incorporated.

\subsection{Detector calibration}

An independent effect to be considered is the calibration of the GW detectors. Currently, the GW parameter estimation results are marginalized over the detector calibration uncertainties ($\lesssim 4\%$ in amplitude {in O1 and O2}), which accounts for both the statistical uncertainty and the systematic error correlated between detections \citep{LIGOScientific:2018mvr}. Both the statistical uncertainty and the systematic error in GW detector calibration are much smaller than the other measurement uncertainties, and thus negligible for $H_0$ estimates from a handful of detections that we have now or expect in the near future \citep{Cahillane2017}. However, the impact of correlated systematic calibration errors between detections will become relatively more important in the long term, with an increasing number of detections driving down the statistical uncertainties, and an improved understanding of other systematic effects that possibly govern our current uncertainty budget. Further quantitative study of the effect of correlated calibration uncertainties is ongoing.

\section{Conclusion and outlook}
\label{sec:conclusion}
In this paper we have presented the first measurement of the Hubble constant
using multiple \ac{GW} observations. Our result reanalyzes and combines the posterior
probability distribution obtained from the \ac{BNS} event GW170817 using the
redshift of the host galaxy inferred from the observed \ac{EM} counterpart
\citep{Abbott:2017xzu}, along with constraints using galaxy catalogs for the
\ac{BBH} events observed by Advanced LIGO and Virgo in their first and second
observing runs. We measure {$H_0=$ {\HubbleMeasCombinedCounterpartFLATLOG} \kmsMpc} ($68.3\%$ highest density posterior interval with a flat-in-log prior). 
This result is mainly dominated by the information from GW170817
with its counterpart, but does show a {modest} improvement with the inclusion of
the \acp{BBH}. The \acp{BBH} contribute both from associated galaxy catalogs as
well as via their observed luminosity distance distribution. 
{Since the latter contribution is sensitive to the assumptions on the mass 
distribution and rate evolution, a more thorough treatment requires a
marginalization over these unknown population parameters. }

The contribution from events
without counterparts is dominated by {GW170814, for which the associated galaxy catalog is highly complete.}
This highlights the importance of deeper surveys and of
dedicated \ac{EM} follow-up of sky regions following \ac{GW} triggers for a
better $H_0$ measurement. With numerous anticipated detections in the upcoming
observing runs with improved detector sensitivities
\citep{Aasi:2013wya,Abbott:2016nhf,TheLIGOScientific:2016pea,TheLIGOScientific:2017qsa,LIGOScientific:2018mvr,LIGOScientific:2018jsj},
these results pave the road towards an era of precision multimessenger
cosmology to be performed with a multitude of sources, including both neutron
star and black hole mergers, with or without transient \ac{EM} counterparts.

\section*{Acknowledgments}
Analyses in this paper made use of
\textsc{NumPy} \citep{Numpy2020},
\textsc{SciPy} \citep{2020SciPy-NMeth},
\textsc{Astropy} \citep{astropy:2013,astropy:2018},
\textsc{gwcosmo} (\url{https://git.ligo.org/lscsoft/gwcosmo}),
and \textsc{IPython} \citep{ipython};
plots were produced with
\textsc{Matplotlib} \citep{matplotlib}, and
\textsc{Seaborn} \citep{seaborn}.
This research made use of the K-corrections calculator service available at \url{http://kcor.sai.msu.ru/}.

The authors gratefully acknowledge the support of the United States
National Science Foundation (NSF) for the construction and operation of the
LIGO Laboratory and Advanced LIGO as well as the Science and Technology Facilities Council (STFC) of the
United Kingdom, the Max-Planck-Society (MPS), and the State of
Niedersachsen/Germany for support of the construction of Advanced LIGO 
and construction and operation of the GEO600 detector. 
Additional support for Advanced LIGO was provided by the Australian Research Council.
The authors gratefully acknowledge the Italian Istituto Nazionale di Fisica Nucleare (INFN),  
the French Centre National de la Recherche Scientifique (CNRS) and
the Foundation for Fundamental Research on Matter supported by the Netherlands Organisation for Scientific Research, 
for the construction and operation of the Virgo detector
and the creation and support  of the EGO consortium. 
The authors also gratefully acknowledge research support from these agencies as well as by 
the Council of Scientific and Industrial Research of India, 
the Department of Science and Technology, India,
the Science \& Engineering Research Board (SERB), India,
the Ministry of Human Resource Development, India,
the Spanish  Agencia Estatal de Investigaci\'on,
the Vicepresid\`encia i Conselleria d'Innovaci\'o, Recerca i Turisme and the Conselleria d'Educaci\'o i Universitat del Govern de les Illes Balears,
the Conselleria d'Educaci\'o, Investigaci\'o, Cultura i Esport de la Generalitat Valenciana,
the National Science Centre of Poland,
the Swiss National Science Foundation (SNSF),
the Russian Foundation for Basic Research, 
the Russian Science Foundation,
the European Commission,
the European Regional Development Funds (ERDF),
the Royal Society, 
the Scottish Funding Council, 
the Scottish Universities Physics Alliance, 
the Hungarian Scientific Research Fund (OTKA),
the Lyon Institute of Origins (LIO),
the Paris \^{I}le-de-France Region, 
the National Research, Development and Innovation Office Hungary (NKFIH), 
the National Research Foundation of Korea,
Industry Canada and the Province of Ontario through the Ministry of Economic Development and Innovation, 
the Natural Science and Engineering Research Council Canada,
the Canadian Institute for Advanced Research,
the Brazilian Ministry of Science, Technology, Innovations, and Communications,
the International Center for Theoretical Physics South American Institute for Fundamental Research (ICTP-SAIFR), 
the Research Grants Council of Hong Kong,
the National Natural Science Foundation of China (NSFC),
the Leverhulme Trust, 
the Research Corporation, 
the Ministry of Science and Technology (MOST), Taiwan
and
the Kavli Foundation.
The authors gratefully acknowledge the support of the NSF, STFC, INFN and CNRS for provision of computational resources.

\bibliographystyle{yahapj}
\bibliography{references}

\begin{thebibliography}{}
\providecommand\natexlab[1]{#1}
\providecommand\JournalTitle[1]{#1}

\bibitem[{Abbott {et~al.}(2016{\natexlab{a}})}]{TheLIGOScientific:2016pea}
Abbott, B.~P., {et~al.} 2016{\natexlab{a}},
  \href{http://dx.doi.org/10.1103/PhysRevX.6.041015}{\JournalTitle{Phys. Rev.
  X}, 6, 041015}, [Erratum: Phys.Rev.X 8, 039903 (2018)]

\bibitem[{Abbott {et~al.}(2016{\natexlab{b}})}]{Abbott:2016nhf}
---. 2016{\natexlab{b}},
  \href{http://dx.doi.org/10.3847/2041-8205/833/1/L1}{\JournalTitle{Astrophys.
  J.}, 833, L1}

\bibitem[{Abbott {et~al.}(2017{\natexlab{a}})}]{Abbott:2017xzu}
---. 2017{\natexlab{a}},
  \href{http://dx.doi.org/10.1038/nature24471}{\JournalTitle{Nature}, 551, 85}

\bibitem[{Abbott {et~al.}(2017{\natexlab{b}})}]{Abbott:2016wiq}
---. 2017{\natexlab{b}},
  \href{http://dx.doi.org/10.1088/1361-6382/aa6854}{\JournalTitle{Class. Quant.
  Grav.}, 34, 104002}

\bibitem[{Abbott {et~al.}(2017{\natexlab{c}})}]{TheLIGOScientific:2017qsa}
---. 2017{\natexlab{c}},
  \href{http://dx.doi.org/10.1103/PhysRevLett.119.161101}{\JournalTitle{Phys.
  Rev. Lett.}, 119, 161101}

\bibitem[{Abbott {et~al.}(2018{\natexlab{a}})}]{Aasi:2013wya}
---. 2018{\natexlab{a}}, \href{http://dx.doi.org/10.1007/s41114-018-0012-9,
  10.1007/lrr-2016-1}{\JournalTitle{Living Rev. Rel.}, 21, 3}

\bibitem[{Abbott {et~al.}(2019{\natexlab{a}})}]{LIGOScientific:2018jsj}
---. 2019{\natexlab{a}},
  \href{http://dx.doi.org/10.3847/2041-8213/ab3800}{\JournalTitle{Astrophys. J.
  Lett.}, 882, L24}

\bibitem[{Abbott {et~al.}(2019{\natexlab{b}})}]{LIGOScientific:2018mvr}
---. 2019{\natexlab{b}},
  \href{http://dx.doi.org/10.1103/PhysRevX.9.031040}{\JournalTitle{Phys. Rev.
  X}, 9, 031040}

\bibitem[{Abbott {et~al.}(2020)}]{Abbott:2020uma}
---. 2020,
  \href{http://dx.doi.org/10.3847/2041-8213/ab75f5}{\JournalTitle{Astrophys. J.
  Lett.}, 892, L3}

\bibitem[{Abbott {et~al.}(2018{\natexlab{b}})}]{Abbott:2018jhe}
Abbott, T. M.~C., {et~al.} 2018{\natexlab{b}},
  \href{http://dx.doi.org/10.3847/1538-4365/aae9f0}{\JournalTitle{Astrophys. J.
  Suppl.}, 239, 18}

\bibitem[{Abolfathi {et~al.}(2018)}]{Abolfathi:2017vfu}
Abolfathi, B., {et~al.} 2018,
  \href{http://dx.doi.org/10.3847/1538-4365/aa9e8a}{\JournalTitle{Astrophys. J.
  Suppl.}, 235, 42}

\bibitem[{Aghanim {et~al.}(2020)}]{Aghanim:2018eyx}
Aghanim, N., {et~al.} 2020,
  \href{http://dx.doi.org/10.1051/0004-6361/201833910}{\JournalTitle{Astron.
  Astrophys.}, 641, A6}

\bibitem[{Ahumada {et~al.}(2020)}]{Ahumada:2019vht}
Ahumada, R., {et~al.} 2020,
  \href{http://dx.doi.org/10.3847/1538-4365/ab929e}{\JournalTitle{Astrophys. J.
  Suppl.}, 249, 3}

\bibitem[{Babak {et~al.}(2017)Babak, Taracchini, \& Buonanno}]{Babak:2016tgq}
Babak, S., Taracchini, A., \& Buonanno, A. 2017,
  \href{http://dx.doi.org/10.1103/PhysRevD.95.024010}{\JournalTitle{Phys.
  Rev.}, D95, 024010}

\bibitem[{Bilicki {et~al.}(2014)Bilicki, Jarrett, Peacock, Cluver, \&
  Steward}]{Bilicki:2013sza}
Bilicki, M., Jarrett, T.~H., Peacock, J.~A., Cluver, M.~E., \& Steward, L.
  2014,
  \href{http://dx.doi.org/10.1088/0067-0049/210/1/9}{\JournalTitle{Astrophys.
  J. Suppl.}, 210, 9}

\bibitem[{Birrer {et~al.}(2019)}]{Birrer:2018vtm}
Birrer, S., {et~al.} 2019,
  \href{http://dx.doi.org/10.1093/mnras/stz200}{\JournalTitle{Mon. Not. Roy.
  Astron. Soc.}, 484, 4726}

\bibitem[{Blanton {et~al.}(2003)Blanton, Hogg, Bahcall, Brinkmann, Britton,
  Connolly, Csabai, Fukugita, Loveday, Meiksin, Munn, Nichol, Okamura, Quinn,
  Schneider, Shimasaku, Strauss, Tegmark, Vogeley, \& Weinberg}]{Blanton_2003}
Blanton, M.~R., Hogg, D.~W., Bahcall, N.~A., {et~al.} 2003,
  \href{http://dx.doi.org/10.1086/375776}{\JournalTitle{The Astrophysical
  Journal}, 592, 819}

\bibitem[{{Caditz} \& {Petrosian}(1989)}]{1989ApJ...337L..65C}
{Caditz}, D., \& {Petrosian}, V. 1989,
  \href{http://dx.doi.org/10.1086/185379}{\JournalTitle{\apjl}, 337, L65}

\bibitem[{Cahillane {et~al.}(2017)Cahillane, Betzwieser, Brown, Goetz, Hall,
  Izumi, Kandhasamy, Karki, Kissel, Mendell, Savage, Tuyenbayev, Urban, Viets,
  Wade, \& Weinstein}]{Cahillane2017}
Cahillane, C., Betzwieser, J., Brown, D.~A., {et~al.} 2017,
  \href{http://dx.doi.org/10.1103/PhysRevD.96.102001}{\JournalTitle{Phys. Rev.
  D}, 96, 102001}

\bibitem[{Carrick {et~al.}(2015)Carrick, Turnbull, Lavaux, \&
  Hudson}]{Carrick:2015xza}
Carrick, J., Turnbull, S.~J., Lavaux, G., \& Hudson, M.~J. 2015,
  \href{http://dx.doi.org/10.1093/mnras/stv547}{\JournalTitle{Mon. Not. Roy.
  Astron. Soc.}, 450, 317}

\bibitem[{Chen {et~al.}(2018)Chen, Fishbach, \& Holz}]{Chen:2017rfc}
Chen, H.-Y., Fishbach, M., \& Holz, D.~E. 2018,
  \href{http://dx.doi.org/10.1038/s41586-018-0606-0}{\JournalTitle{Nature},
  562, 545}

\bibitem[{Chen \& Holz(2014)}]{Chen:2014yla}
Chen, H.-Y., \& Holz, D.~E. 2014,
  \href{http://arxiv.org/abs/1409.0522}{{\sffamily arXiv:1409.0522 [gr-qc]}}

\bibitem[{Chilingarian {et~al.}(2010)Chilingarian, Melchior, \&
  Zolotukhin}]{Kcorcalc2010}
Chilingarian, I.~V., Melchior, A.-L., \& Zolotukhin, I.~Y. 2010,
  \href{http://dx.doi.org/10.1111/j.1365-2966.2010.16506.x}{\JournalTitle{Monthly
  Notices of the Royal Astronomical Society}, 405, 1409}

\bibitem[{{Chilingarian} \& {Zolotukhin}(2012)}]{Kcorcalc2012}
{Chilingarian}, I.~V., \& {Zolotukhin}, I.~Y. 2012,
  \href{http://dx.doi.org/10.1111/j.1365-2966.2011.19837.x}{\JournalTitle{\mnras},
  419, 1727}

\bibitem[{Dalal {et~al.}(2006)Dalal, Holz, Hughes, \& Jain}]{Dalal:2006qt}
Dalal, N., Holz, D.~E., Hughes, S.~A., \& Jain, B. 2006,
  \href{http://dx.doi.org/10.1103/PhysRevD.74.063006}{\JournalTitle{Phys.
  Rev.}, D74, 063006}

\bibitem[{Del~Pozzo(2012)}]{DelPozzo:2011yh}
Del~Pozzo, W. 2012,
  \href{http://dx.doi.org/10.1103/PhysRevD.86.043011}{\JournalTitle{Phys.
  Rev.}, D86, 043011}

\bibitem[{Drlica-Wagner {et~al.}(2018)}]{Drlica-Wagner:2017tkk}
Drlica-Wagner, A., {et~al.} 2018,
  \href{http://dx.doi.org/10.3847/1538-4365/aab4f5}{\JournalTitle{Astrophys. J.
  Suppl.}, 235, 33}

\bibitem[{Dálya {et~al.}(2018)Dálya, Galgóczi, Dobos, Frei, Heng, Macas,
  Messenger, Raffai, \& de~Souza}]{Dalya:2018cnd}
Dálya, G., Galgóczi, G., Dobos, L., {et~al.} 2018,
  \href{http://dx.doi.org/10.1093/mnras/sty1703}{\JournalTitle{Mon. Not. Roy.
  Astron. Soc.}, 479, 2374}

\bibitem[{Farr {et~al.}(2019)Farr, Fishbach, Ye, \& Holz}]{Farr:2019twy}
Farr, W.~M., Fishbach, M., Ye, J., \& Holz, D. 2019,
  \href{http://dx.doi.org/10.3847/2041-8213/ab4284}{\JournalTitle{Astrophys.
  J.}, 883, L42}

\bibitem[{Feeney {et~al.}(2019)Feeney, Peiris, Williamson, Nissanke, Mortlock,
  Alsing, \& Scolnic}]{Feeney:2018mkj}
Feeney, S.~M., Peiris, H.~V., Williamson, A.~R., {et~al.} 2019,
  \href{http://dx.doi.org/10.1103/PhysRevLett.122.061105}{\JournalTitle{Phys.
  Rev. Lett.}, 122, 061105}

\bibitem[{Finn(1994)}]{Finn:1994cg}
Finn, L.~S. 1994,
  \href{http://dx.doi.org/10.1103/PhysRevLett.73.1878}{\JournalTitle{Phys. Rev.
  Lett.}, 73, 1878}

\bibitem[{Fishbach {et~al.}(2019)Fishbach, Gray, Magaña~Hernandez, Qi, \&
  Sur}]{Fishbach:2018gjp}
Fishbach, M., Gray, R., Magaña~Hernandez, I., Qi, H., \& Sur, A. 2019,
  \href{http://dx.doi.org/10.3847/2041-8213/aaf96e}{\JournalTitle{Astrophys.
  J.}, 871, L13}

\bibitem[{Fishbach {et~al.}(2018)Fishbach, Holz, \& Farr}]{Fishbach:2018edt}
Fishbach, M., Holz, D.~E., \& Farr, W.~M. 2018,
  \href{http://dx.doi.org/10.3847/2041-8213/aad800}{\JournalTitle{Astrophys.
  J.}, 863, L41}

\bibitem[{Fowler \& Hoyle(1964)}]{Fowler:1964zz}
Fowler, W.~A., \& Hoyle, F. 1964,
  \href{http://dx.doi.org/10.1086/190103}{\JournalTitle{Astrophys. J. Suppl.},
  9, 201}

\bibitem[{Gehrels {et~al.}(2016)Gehrels, Cannizzo, Kanner, Kasliwal, Nissanke,
  \& Singer}]{Gehrels:2015uga}
Gehrels, N., Cannizzo, J.~K., Kanner, J., {et~al.} 2016,
  \href{http://dx.doi.org/10.3847/0004-637X/820/2/136}{\JournalTitle{Astrophys.
  J.}, 820, 136}

\bibitem[{Gray {et~al.}(2020)}]{Gray:2019ksv}
Gray, R., {et~al.} 2020,
  \href{http://dx.doi.org/10.1103/PhysRevD.101.122001}{\JournalTitle{Phys. Rev.
  D}, 101, 122001}

\bibitem[{Hannam {et~al.}(2014)Hannam, Schmidt, Bohé, Haegel, Husa, Ohme,
  Pratten, \& Pürrer}]{Hannam:2013oca}
Hannam, M., Schmidt, P., Bohé, A., {et~al.} 2014,
  \href{http://dx.doi.org/10.1103/PhysRevLett.113.151101}{\JournalTitle{Phys.
  Rev. Lett.}, 113, 151101}

\bibitem[{Harris {et~al.}(2020)Harris, Millman, van~der Walt, Gommers,
  Virtanen, Cournapeau, Wieser, Taylor, Berg, Smith, Kern, Picus, Hoyer, van
  Kerkwijk, Brett, Haldane, del R{\'\i}o, Wiebe, Peterson, G{\'e}rard-Marchant,
  Sheppard, Reddy, Weckesser, Abbasi, Gohlke, \& Oliphant}]{Numpy2020}
Harris, C.~R., Millman, K.~J., van~der Walt, S.~J., {et~al.} 2020,
  \href{http://dx.doi.org/10.1038/s41586-020-2649-2}{\JournalTitle{Nature},
  585, 357}

\bibitem[{Hogg {et~al.}(2002)Hogg, Baldry, Blanton, \&
  Eisenstein}]{Hogg:2002yh}
Hogg, D.~W., Baldry, I.~K., Blanton, M.~R., \& Eisenstein, D.~J. 2002,
  \href{http://arxiv.org/abs/astro-ph/0210394}{{\sffamily
  arXiv:astro-ph/0210394}}

\bibitem[{Holz \& Hughes(2005)}]{Holz:2005df}
Holz, D.~E., \& Hughes, S.~A. 2005,
  \href{http://dx.doi.org/10.1086/431341}{\JournalTitle{Astrophys. J.}, 629,
  15}

\bibitem[{Hoyle {et~al.}(2018)}]{Hoyle:2017mee}
Hoyle, B., {et~al.} 2018,
  \href{http://dx.doi.org/10.1093/mnras/sty957}{\JournalTitle{Mon. Not. Roy.
  Astron. Soc.}, 478, 592}

\bibitem[{{Hunter}(2007)}]{matplotlib}
{Hunter}, J.~D. 2007,
  \href{http://dx.doi.org/10.1109/MCSE.2007.55}{\JournalTitle{Computing in
  Science Engineering}, 9, 90}

\bibitem[{Husa {et~al.}(2016)Husa, Khan, Hannam, Pürrer, Ohme,
  Jiménez~Forteza, \& Bohé}]{Husa:2015iqa}
Husa, S., Khan, S., Hannam, M., {et~al.} 2016,
  \href{http://dx.doi.org/10.1103/PhysRevD.93.044006}{\JournalTitle{Phys.
  Rev.}, D93, 044006}

\bibitem[{Khan {et~al.}(2016)Khan, Husa, Hannam, Ohme, Pürrer,
  Jiménez~Forteza, \& Bohé}]{Khan:2015jqa}
Khan, S., Husa, S., Hannam, M., {et~al.} 2016,
  \href{http://dx.doi.org/10.1103/PhysRevD.93.044007}{\JournalTitle{Phys.
  Rev.}, D93, 044007}

\bibitem[{Kiziltan {et~al.}(2013)Kiziltan, Kottas, De~Yoreo, \&
  Thorsett}]{Kiziltan:2013oja}
Kiziltan, B., Kottas, A., De~Yoreo, M., \& Thorsett, S.~E. 2013,
  \href{http://dx.doi.org/10.1088/0004-637X/778/1/66}{\JournalTitle{Astrophys.
  J.}, 778, 66}

\bibitem[{Macaulay {et~al.}(2019)}]{Macaulay:2018fxi}
Macaulay, E., {et~al.} 2019,
  \href{http://dx.doi.org/10.1093/mnras/stz978}{\JournalTitle{Mon. Not. Roy.
  Astron. Soc.}, 486, 2184}

\bibitem[{MacLeod \& Hogan(2008)}]{MacLeod:2007jd}
MacLeod, C.~L., \& Hogan, C.~J. 2008,
  \href{http://dx.doi.org/10.1103/PhysRevD.77.043512}{\JournalTitle{Phys.
  Rev.}, D77, 043512}

\bibitem[{{Madgwick} {et~al.}(2002){Madgwick}, {Lahav}, {Baldry}, {Baugh},
  {Bland-Hawthorn}, {Bridges}, {Cannon}, {Cole}, {Colless}, {Collins}, {Couch},
  {Dalton}, {De Propris}, {Driver}, {Efstathiou}, {Ellis}, {Frenk},
  {Glazebrook}, {Jackson}, {Lewis}, {Lumsden}, {Maddox}, {Norberg}, {Peacock},
  {Peterson}, {Sutherland}, \& {Taylor}}]{2002MNRAS.333..133M}
{Madgwick}, D.~S., {Lahav}, O., {Baldry}, I.~K., {et~al.} 2002,
  \href{http://dx.doi.org/10.1046/j.1365-8711.2002.05393.x}{\JournalTitle{\mnras},
  333, 133}

\bibitem[{Makarov {et~al.}(2014)Makarov, Prugniel, Terekhova, Courtois, \&
  Vauglin}]{Makarov:2014txa}
Makarov, D., Prugniel, P., Terekhova, N., Courtois, H., \& Vauglin, I. 2014,
  \href{http://dx.doi.org/10.1051/0004-6361/201423496}{\JournalTitle{Astron.
  Astrophys.}, 570, A13}

\bibitem[{Mandel \& de~Mink(2016)}]{Mandel:2015qlu}
Mandel, I., \& de~Mink, S.~E. 2016,
  \href{http://dx.doi.org/10.1093/mnras/stw379}{\JournalTitle{Mon. Not. Roy.
  Astron. Soc.}, 458, 2634}

\bibitem[{Mandel {et~al.}(2019)Mandel, Farr, \& Gair}]{Mandel:2018mve}
Mandel, I., Farr, W.~M., \& Gair, J.~R. 2019,
  \href{http://dx.doi.org/10.1093/mnras/stz896}{\JournalTitle{Mon. Not. Roy.
  Astron. Soc.}, 486, 1086}

\bibitem[{Messenger \& Read(2012)}]{Messenger:2011gi}
Messenger, C., \& Read, J. 2012,
  \href{http://dx.doi.org/10.1103/PhysRevLett.108.091101}{\JournalTitle{Phys.
  Rev. Lett.}, 108, 091101}

\bibitem[{Mortlock {et~al.}(2019)Mortlock, Feeney, Peiris, Williamson, \&
  Nissanke}]{Mortlock:2018azx}
Mortlock, D.~J., Feeney, S.~M., Peiris, H.~V., Williamson, A.~R., \& Nissanke,
  S.~M. 2019,
  \href{http://dx.doi.org/10.1103/PhysRevD.100.103523}{\JournalTitle{Phys. Rev.
  D}, 100, 103523}

\bibitem[{Nair {et~al.}(2018)Nair, Bose, \& Saini}]{Nair:2018ign}
Nair, R., Bose, S., \& Saini, T.~D. 2018,
  \href{http://dx.doi.org/10.1103/PhysRevD.98.023502}{\JournalTitle{Phys.
  Rev.}, D98, 023502}

\bibitem[{Nissanke {et~al.}(2013)Nissanke, Holz, Dalal, Hughes, Sievers, \&
  Hirata}]{Nissanke:2013fka}
Nissanke, S., Holz, D.~E., Dalal, N., {et~al.} 2013,
  \href{http://arxiv.org/abs/1307.2638}{{\sffamily arXiv:1307.2638
  [astro-ph.CO]}}

\bibitem[{Nissanke {et~al.}(2010)Nissanke, Holz, Hughes, Dalal, \&
  Sievers}]{Nissanke:2009kt}
Nissanke, S., Holz, D.~E., Hughes, S.~A., Dalal, N., \& Sievers, J.~L. 2010,
  \href{http://dx.doi.org/10.1088/0004-637X/725/1/496}{\JournalTitle{Astrophys.
  J.}, 725, 496}

\bibitem[{{Oke} \& {Sandage}(1968)}]{Oke1968}
{Oke}, J.~B., \& {Sandage}, A. 1968,
  \href{http://dx.doi.org/10.1086/149737}{\JournalTitle{\apj}, 154, 21}

\bibitem[{Pan {et~al.}(2014)Pan, Buonanno, Taracchini, Kidder, Mroué,
  Pfeiffer, Scheel, \& Szilágyi}]{Pan:2013rra}
Pan, Y., Buonanno, A., Taracchini, A., {et~al.} 2014,
  \href{http://dx.doi.org/10.1103/PhysRevD.89.084006}{\JournalTitle{Phys.
  Rev.}, D89, 084006}

\bibitem[{{Perez} \& {Granger}(2007)}]{ipython}
{Perez}, F., \& {Granger}, B.~E. 2007,
  \href{http://dx.doi.org/10.1109/MCSE.2007.53}{\JournalTitle{Computing in
  Science Engineering}, 9, 21}

\bibitem[{{Planck Collaboration}(2016)}]{Ade:2015xua}
{Planck Collaboration}. 2016,
  \href{http://dx.doi.org/10.1051/0004-6361/201525830}{\JournalTitle{Astron.
  Astrophys.}, 594, A13}

\bibitem[{Price-Whelan {et~al.}(2018)}]{astropy:2018}
Price-Whelan, A., {et~al.} 2018,
  \href{http://dx.doi.org/10.3847/1538-3881/aabc4f}{\JournalTitle{Astron. J.},
  156, 123}

\bibitem[{Pâris {et~al.}(2017)}]{Paris:2016xdm}
Pâris, I., {et~al.} 2017,
  \href{http://dx.doi.org/10.1051/0004-6361/201527999}{\JournalTitle{Astron.
  Astrophys.}, 597, A79}

\bibitem[{Rahman {et~al.}(2019)Rahman, Nissanke, Williamson, Schmidt, Singer,
  Hirata, \& Gehrels}]{GWENS}
Rahman, M., Nissanke, S., Williamson, A., {et~al.} 2019, \JournalTitle{(in
  preparation)}

\bibitem[{Riess {et~al.}(2019)Riess, Casertano, Yuan, Macri, \&
  Scolnic}]{Riess:2019cxk}
Riess, A.~G., Casertano, S., Yuan, W., Macri, L.~M., \& Scolnic, D. 2019,
  \href{http://dx.doi.org/10.3847/1538-4357/ab1422}{\JournalTitle{Astrophys.
  J.}, 876, 85}

\bibitem[{Robitaille {et~al.}(2013)}]{astropy:2013}
Robitaille, T.~P., {et~al.} 2013,
  \href{http://dx.doi.org/10.1051/0004-6361/201322068}{\JournalTitle{Astron.
  Astrophys.}, 558, A33}

\bibitem[{Sadeh {et~al.}(2016)Sadeh, Abdalla, \& Lahav}]{Sadeh:2015lsa}
Sadeh, I., Abdalla, F.~B., \& Lahav, O. 2016,
  \href{http://dx.doi.org/10.1088/1538-3873/128/968/104502}{\JournalTitle{Publ.
  Astron. Soc. Pac.}, 128, 104502}

\bibitem[{Sathyaprakash {et~al.}(2010)Sathyaprakash, Schutz, \& Van
  Den~Broeck}]{Sathyaprakash:2009xt}
Sathyaprakash, B.~S., Schutz, B.~F., \& Van Den~Broeck, C. 2010,
  \href{http://dx.doi.org/10.1088/0264-9381/27/21/215006}{\JournalTitle{Class.
  Quant. Grav.}, 27, 215006}

\bibitem[{Saunders {et~al.}(1990)Saunders, Rowan-Robinson, Lawrence,
  Efstathiou, Kaiser, Ellis, \& Frenk}]{10.1093/mnras/242.3.318}
Saunders, W., Rowan-Robinson, M., Lawrence, A., {et~al.} 1990,
  \href{http://dx.doi.org/10.1093/mnras/242.3.318}{\JournalTitle{Monthly
  Notices of the Royal Astronomical Society}, 242, 318}

\bibitem[{Schutz(1986)}]{Schutz:1986gp}
Schutz, B.~F. 1986,
  \href{http://dx.doi.org/10.1038/323310a0}{\JournalTitle{Nature}, 323, 310}

\bibitem[{Sevilla-Noarbe {et~al.}(2018)}]{Sevilla-Noarbe:2018ktd}
Sevilla-Noarbe, I., {et~al.} 2018,
  \href{http://dx.doi.org/10.1093/mnras/sty2579}{\JournalTitle{Mon. Not. Roy.
  Astron. Soc.}, 481, 5451}

\bibitem[{Skrutskie {et~al.}(2006)}]{Skrutskie:2006wh}
Skrutskie, M.~F., {et~al.} 2006,
  \href{http://dx.doi.org/10.1086/498708}{\JournalTitle{Astron. J.}, 131, 1163}

\bibitem[{Soares-Santos {et~al.}(2019)Soares-Santos, Palmese,
  {et~al.}}]{Soares-Santos:2019irc}
Soares-Santos, M., Palmese, A., {et~al.} 2019,
  \href{http://dx.doi.org/10.3847/2041-8213/ab14f1}{\JournalTitle{Astrophys.
  J.}, 876, L7}

\bibitem[{Taracchini {et~al.}(2014)}]{Taracchini:2013rva}
Taracchini, A., {et~al.} 2014,
  \href{http://dx.doi.org/10.1103/PhysRevD.89.061502}{\JournalTitle{Phys.
  Rev.}, D89, 061502}

\bibitem[{Taylor \& Gair(2012)}]{Taylor:2012db}
Taylor, S.~R., \& Gair, J.~R. 2012,
  \href{http://dx.doi.org/10.1103/PhysRevD.86.023502}{\JournalTitle{Phys.
  Rev.}, D86, 023502}

\bibitem[{Taylor {et~al.}(2012)Taylor, Gair, \& Mandel}]{Taylor:2011fs}
Taylor, S.~R., Gair, J.~R., \& Mandel, I. 2012,
  \href{http://dx.doi.org/10.1103/PhysRevD.85.023535}{\JournalTitle{Phys. Rev.
  D}, 85, 023535}

\bibitem[{{Virtanen} {et~al.}(2020){Virtanen}, {Gommers}, {Oliphant},
  {Haberland}, {Reddy}, {Cournapeau}, {Burovski}, {Peterson}, {Weckesser},
  {Bright}, {van der Walt}, {Brett}, {Wilson}, {Jarrod Millman}, {Mayorov},
  {Nelson}, {Jones}, {Kern}, {Larson}, {Carey}, {Polat}, {Feng}, {Moore}, {Vand
  erPlas}, {Laxalde}, {Perktold}, {Cimrman}, {Henriksen}, {Quintero}, {Harris},
  {Archibald}, {Ribeiro}, {Pedregosa}, {van Mulbregt}, \&
  {Contributors}}]{2020SciPy-NMeth}
{Virtanen}, P., {Gommers}, R., {Oliphant}, T.~E., {et~al.} 2020,
  \href{http://dx.doi.org/https://doi.org/10.1038/s41592-019-0686-2}{\JournalTitle{Nature
  Methods}, 17, 261}

\bibitem[{Vitale \& Chen(2018)}]{Vitale:2018wlg}
Vitale, S., \& Chen, H.-Y. 2018,
  \href{http://dx.doi.org/10.1103/PhysRevLett.121.021303}{\JournalTitle{Phys.
  Rev. Lett.}, 121, 021303}

\bibitem[{Vitale {et~al.}(2020)Vitale, Gerosa, Farr, \&
  Taylor}]{Vitale:2020aaz}
Vitale, S., Gerosa, D., Farr, W.~M., \& Taylor, S.~R. 2020,
  \href{http://arxiv.org/abs/2007.05579}{{\sffamily arXiv:2007.05579
  [astro-ph.IM]}}

\bibitem[{Waskom {et~al.}(2020)Waskom, Botvinnik, Ostblom, Gelbart, Lukauskas,
  Hobson, Gemperline, Augspurger, Halchenko, Cole, Warmenhoven, de~Ruiter, Pye,
  Hoyer, Vanderplas, Villalba, Kunter, Quintero, Bachant, Martin, Meyer, Swain,
  Miles, Brunner, O'Kane, Yarkoni, Williams, Evans, Fitzgerald, \&
  Brian}]{seaborn}
Waskom, M., Botvinnik, O., Ostblom, J., {et~al.} 2020, mwaskom/seaborn: v0.10.1
  (April 2020)

\bibitem[{White {et~al.}(2011)White, Daw, \& Dhillon}]{White:2011qf}
White, D.~J., Daw, E.~J., \& Dhillon, V.~S. 2011,
  \href{http://dx.doi.org/10.1088/0264-9381/28/8/085016}{\JournalTitle{Class.
  Quant. Grav.}, 28, 085016}

\end{thebibliography}

\end{document}